\documentclass[twocolumn]{elsart3p}

\usepackage{graphicx,subfig}
\usepackage{epsfig}
\usepackage{amsmath,amssymb,wasysym}

\usepackage{ifthen}
\newcounter{minutesToday}
\newcounter{hours}
\newcounter{hmins}
\newcounter{minutes}
\divide \value{hours} by 60
\multiply \value{hmins} by 60
\def\HHMM{\ifthenelse{\theminutes<10}{\thehours:0\theminutes}{\thehours:\theminutes}}

\usepackage{array,url}
\usepackage{color}

\reversemarginpar

\newcommand{\degree}{\ensuremath{^\circ}}

\def\PermaFlex{PermaFlex\,}
\def\Tyvek{Tyvek\textsuperscript{\textregistered}\,}

\def\Viton{Viton\textsuperscript{\textregistered}\,}
\def\erf{\mbox{erf}}

\begin{document}

\begin{frontmatter}

  \title{The muon system of the Daya Bay Reactor antineutrino experiment }

\def\IHEP{1}
\def\ECUST{2}
\def\UW{3}
\def\BNL{4}
\def\NUU{5}
\def\Dubna{6}
\def\CalTech{7}
\def\CUHK{8}
\def\NCTU{9}
\def\NJU{10}
\def\Tsinghua{11}
\def\SZU{12}
\def\NCEPU{13}
\def\Siena{14}
\def\IIT{15}
\def\LBNL{16}
\def\UCB{17}
\def\UIUC{18}
\def\CDUT{19}
\def\SJTU{20}
\def\BNU{21}
\def\WM{22}
\def\Yale{23}
\def\VT{24}
\def\NTU{25}
\def\CIAE{26}
\def\UCLA{27}
\def\SDU{28}
\def\NanKai{29}
\def\UC{30}
\def\DGUT{31}
\def\HKU{32}
\def\UH{33}
\def\Charles{34}
\def\USTC{35}
\def\SYSU{36}
\def\Princeton{37}
\def\RPI{38}
\def\Temple{39}
\def\CGNPG{40}
\def\NDUT{41}
\def\ISU{42}
\def\XJTU{43}
\def\CUC{44}
\author[\ECUST,\IHEP]{F.P.~An}
\author[\UW]{A.B.~Balantekin}
\author[\Yale,\UW]{H.R.~Band}
\author[\BNL]{W.~Beriguete}
\author[\BNL]{M.~Bishai}
\author[\NTU,\NUU]{S.~Blyth}
\author[\BNL]{R.E.~Brown}
\author[\Dubna]{I.~Butorov}
\author[\IHEP]{G.F.~Cao}
\author[\IHEP]{J.~Cao}
% ...
\setlength{\unitlength}{1cm}

\vspace{-0.2in}

  \begin{abstract}
The Daya Bay experiment consists of functionally identical antineutrino detectors immersed in pools of ultrapure water
in three well-separated underground experimental halls near two nuclear reactor complexes.
These pools serve both as shields against natural, low-energy radiation, and
as water Cherenkov detectors that efficiently detect cosmic muons using arrays of photomultiplier tubes.
Each pool is covered by a plane of resistive plate chambers as an additional means of detecting muons.
Design, construction, operation, and performance of these muon detectors are described.
  \end{abstract}
  

\end{frontmatter}

\section{Introduction}

\begin{figure}
  \begin{center}
    \includegraphics[width=7.8cm]{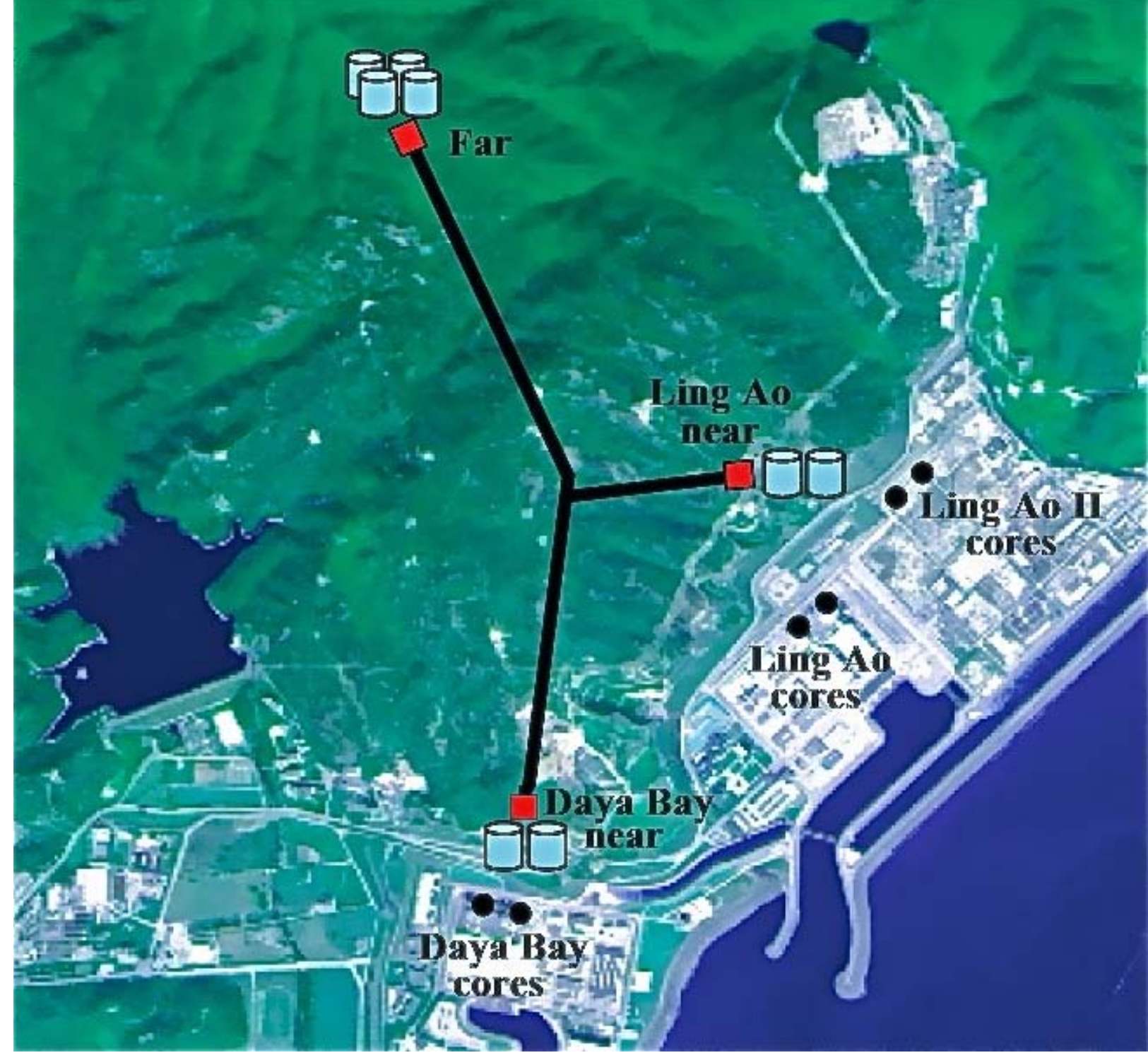}
    \caption{Layout of the experimental halls and reactor cores.
             The reactor cores are indicated by black circles.
             The near experimental halls are the Daya Bay hall (EH1) and the Ling Ao hall (EH2). The far hall is EH3. }
    \label{fig1}
  \end{center}
\end{figure}

The Daya Bay Reactor Antineutrino Experiment is designed to
determine the last unknown neutrino mixing angle $\theta_{13}$ by
observing antineutrino oscillations,
with the quantity $\sin^2 2 \theta_{13}$ measured to a precision of $0.01$ or smaller~\cite{hep-ex/0701029}.
The heart of the Daya Bay experiment is its set of eight functionally identical antineutrino detectors (ADs)~\cite{ADpaper},
distributed amongst three experimental halls located underground to suppress the cosmic muon flux.
Six reactor cores at the nearby Daya Bay and Ling Ao reactor complexes provide a total thermal power of 17.4~GW.
Two experimental halls close to the reactor cores
measure the mostly unoscillated antineutrino spectrum,
while a third experimental hall at a baseline of about 2~km measures the spectrum near the $\theta_{13}$ oscillation maximum
(Fig. \ref{fig1}).
From surveys of the mountain profile and granite cores,
which determined the bulk density to be about 2600~kg/m$^3$, 
the overburden is 860 meters-water-equivalent (mwe) at the far hall, and
250 mwe and 265 mwe at the near halls.

Antineutrinos are detected in the ADs via the inverse beta decay (IBD) interaction~\cite{IBD_CR,IBD_RC}
$\bar{\nu}_{e} p \rightarrow e^{+} n$.
The IBD signature is a coincidence between the prompt energy deposit from the positron and
the delayed release of an 8~MeV $\gamma$ cascade from neutron capture on gadolinium~\cite{1stOsc,2ndOsc}.
The delay occurs while the neutron thermalizes, with a mean time to capture of about 30~$\mu$s.

The {\em raison d'\^etre\/} for the Daya Bay Muon System is to prevent (as a passive shield)
and to eliminate (as an offline veto)
nearly all IBD candidates that are not actually reactor antineutrino captures in an AD.
To this end, the ADs in each hall are immersed in a pool of ultrapure water which
shields the ADs from radioactivity of the surrounding rock and other materials.
The pools are instrumented with photomultiplier tubes (PMTs) to serve as water Cherenkov detectors,
tagging muons that can produce cosmogenic backgrounds such as fast neutrons, ${}^9$Li, and ${}^8$He.

\begin{figure}
\begin{center}
    \includegraphics[width=7.8cm]{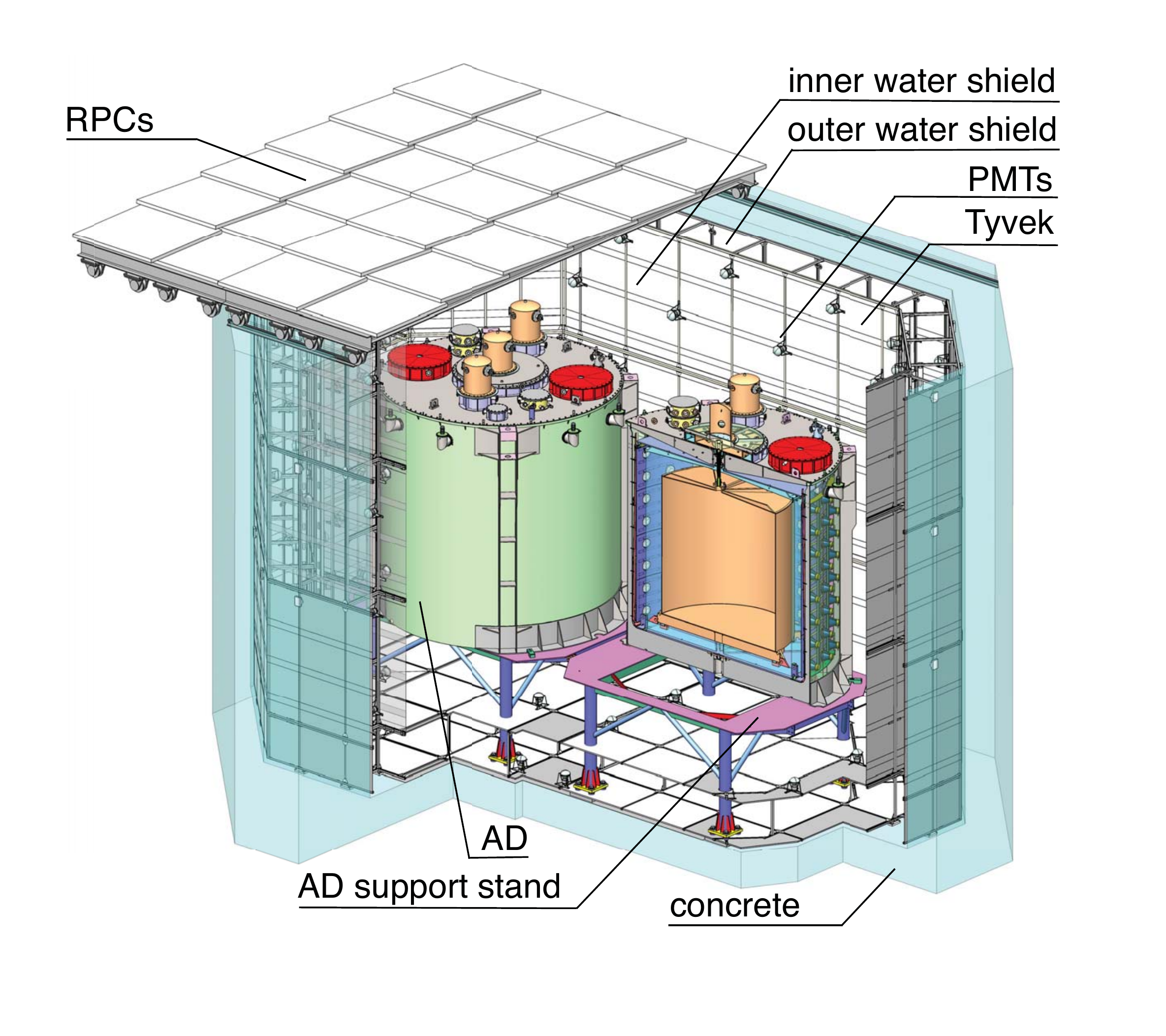}
    \caption{Near hall configuration showing two ADs in a pool, one in cross section.
             The near hall pools are 16~m long by 10~m wide.
             The far hall pool houses four ADs and is 16~m long by 16~m wide.
             The pools are all 10~m deep.
             An optical barrier made of Tyvek divides the pool into inner and outer water shields. }
    \label{fig2}
\end{center}
\end{figure}

Each near hall houses two ADs (Fig. \ref{fig2}), and the far hall houses four.
A \Tyvek optical barrier divides each pool into inner and outer water shields (IWS \& OWS),
both populated with PMTs as shown in Table~\ref{Tab1}.
The OWS comprises the outer $1$~m of each pool's sides and bottom
(but not the top -- the IWS extends all the way to the surface).
At least $2.5$~m of water shields each AD from every direction,
reducing the expected AD PMT rate from rock radioactivity by a factor of about $10^6$, to less than 50~Hz in each hall.
The expected neutron rate per day (summed over all ADs in a hall) is 18 in EH1, 12 in EH2, and 1.5 in EH3.

\begin{table}
  \centering
  \caption{ Muon PMT population.
            The IWS PMTs all face inward (towards the ADs),
            while some OWS PMTs face outward.
          }
  \label{Tab1}
  \vspace{5pt}
  \begin{tabular*}{7cm}{l@{\extracolsep\fill}ccc}
     \hline
     \hline
     Hall & {\sf IWS} & {\sf OWS (inward/outward)} & {\sf Total} \\
     \hline
     EH1          & 121        &   167 (103/64) & 288 \\
     EH2          & 121        &   167 (103/64) & 288 \\
     EH3          & 160        &   224 (128/96) & 384 \\
     \hline
     \hline
   \end{tabular*}
   \centering
\end{table}
An array of RPCs covers each pool, extending about 1~m beyond the pool on each side.
The RPCs detect about one third of the muons which 
miss the pool but which are still close enough to contribute fast neutron and other cosmogenic backgrounds to the IBD signal.

\section{The water shield and muon detector}

\subsection{Water}
\label{SubWater}
The Daya Bay design specifies water with an attenuation length of at least $30$~m for wavelengths near 420~nm
and less than 5~Bq/m${}^3$ of radon.
The water system consists of a dedicated polishing station in each of the three halls,
which delivers ultrapure water to its pool at a resistivity of 18~M$\Omega$-cm and less than 10~ppb O${}_2$,
at a flow rate of 5~m${}^3$/hr at the near halls (small pools)
and 8~m${}^3$/hr at the far hall (large pool)~\cite{fullWater}.
Located centrally to the three experimental halls is a station that
pre-treats water from the civil water supply and feeds the three halls during filling and as needed after filling to
maintain constant levels in the pools.

A comprehensive program was developed to test all candidate materials for
compatibility with ultrapure water, to ensure that nothing that would
significantly degrade the water was allowed into the pools.
As part of this program,
a small prototype detector was constructed at
the Institute for High Energy Physics, Beijing (IHEP) with a circulation and purification system
capable of producing water with an attenuation length of 80~m near 420~nm~\cite{IHEPprototype}.
This prototype system guided the design of the water system for Daya Bay.

The sub-tropical climate at Daya Bay required the water system to cool the water.
This need was confirmed by measurements of the underground rock temperature (27-28~C),
and reports that the water source was a shallow surface pool which,
in the long Daya Bay summer, could reach temperatures of 35~C and perhaps even warmer.
As a compromise between lower noise rates in the AD PMTs for cooler temperatures
and the cost of greater cooling capacity,
the cooling capacity was designed to maintain a water temperature of 24~C,
maintained to within less than 1~C.

\subsection{Photomultiplier Tubes}
The pools are instrumented with two types of pressure-resistant 20~cm hemispherical PMTs,
powered by CAEN
48-channel A1932AP modules housed in SY 1527LC mainframes.

Newly purchased from Hamamatsu, 619 PMTs are 10-stage 20~cm model R5912,
complete with waterproof base assemblies built by Hamamatsu to our specifications.
These PMTs are rated to withstand a pressure of up to seven atmospheres,
over three times the pressure at the bottom of a pool.
The applied positive high voltage and signal are carried with a single 
52-m-long Belden YR-29304 50$\,\Omega$ coaxial cable terminated 
with a Huber$\verb!+!$Suhner 11 SHV-50-4-1 connector. 
The outer jacket of the coaxial cable is high density polyethylene,
compatible with ultra-pure water.

Recycled from the MACRO experiment~\cite{macro},
341 PMTs are 8" EMI
14-stage models 9350KA and D642KB.
These were assembled with custom bases and waterproofed. 
A 52-m-long JUDD C07947 50$\,\Omega$ coaxial cable with a polyethylene outer jacket 
handles the supplied high voltage and signal. 
The EMI PMT assemblies were all tested in a pressure vessel filled with water to about 85~kPa~(gauge).

Prior to installation,
the dark rate, rise time, linearity,  pre- and after-pulsing probabilities,
the peak-to-valley ratio of the digitized single photoelectron waveform,  
the gain as a function of applied voltage,
and the relative detection efficiency of each PMT assembly were measured and archived.
At a gain of $2 \times 10^7$,
the PMTs demonstrate a single photoelectron (pe) peak-to-valley ratio of $2.5$ or better.
A small percentage of the assemblies tested was rejected,
such as those with dark rates exceeding $10$~kHz at a threshold of 0.25~pe.
The characterization of each PMT determined its HV setting for a nominal data-taking gain of $1.0 \times 10^7$,
about 20 ADC counts per photoelectron.

\begin{figure}
\begin{center}
  \includegraphics[width=7.0cm]{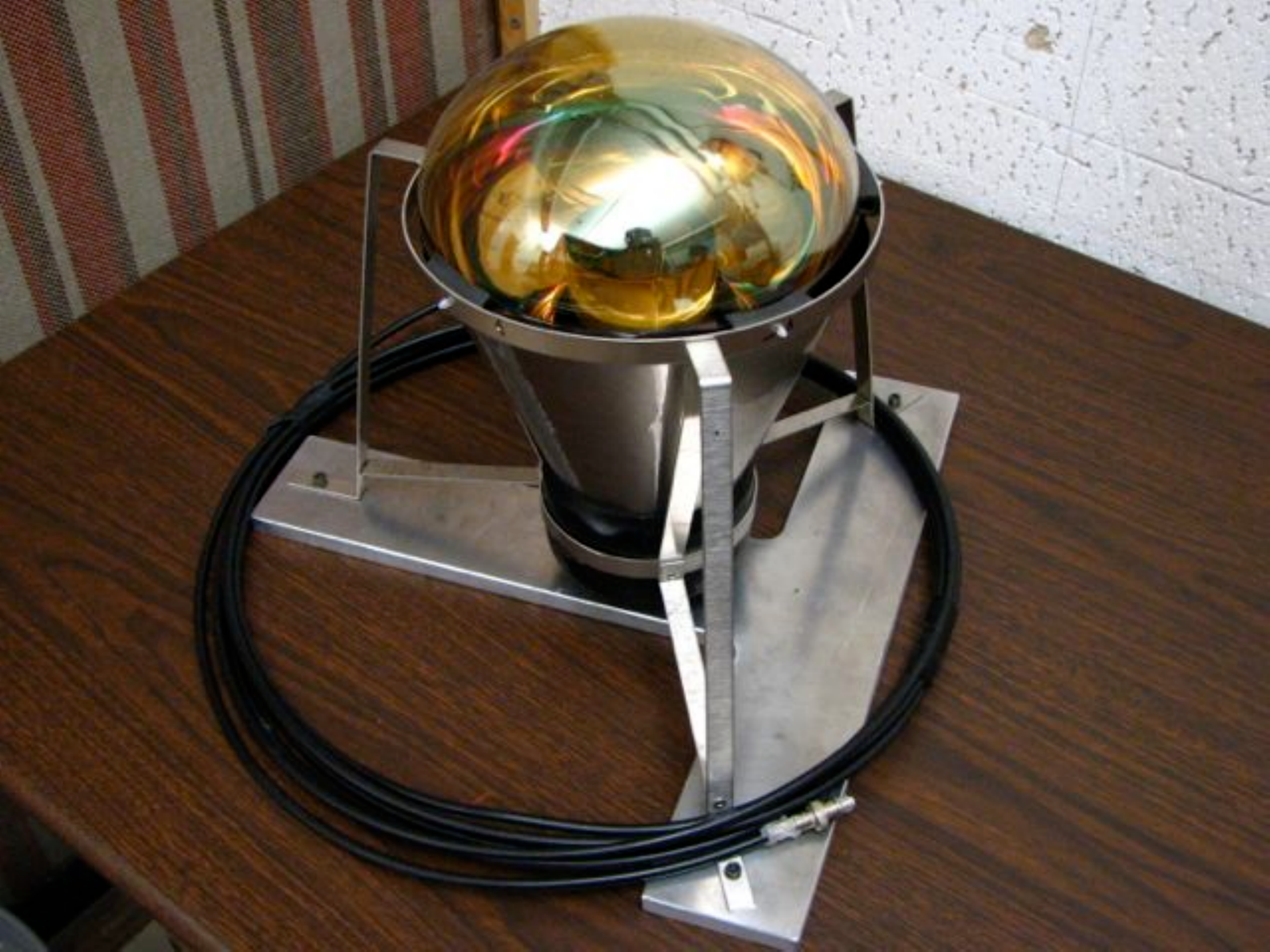}
  \caption{ Photograph of a Muon PMT assembly, including PMT, base, cable, magnetic shield, bracket, and ``Tee''. }
  \label{Fig3}
\end{center}
\end{figure}

The PMTs were each pre-assembled with a magnetic shield~\cite{MagShield}, a bracket, and ``Tee'' support
(Fig. \ref{Fig3}).
The brackets and ``Tees'' are type-304 stainless steel.
Small pieces of \Viton prevent direct contact between the brackets and PMTs.
The PMT assemblies were carefully boxed and labeled with pre-determined installation positions.

\subsection{Resistive Plate Chambers}

\begin{figure}
  \begin{center}
    \includegraphics[width=7.8cm]{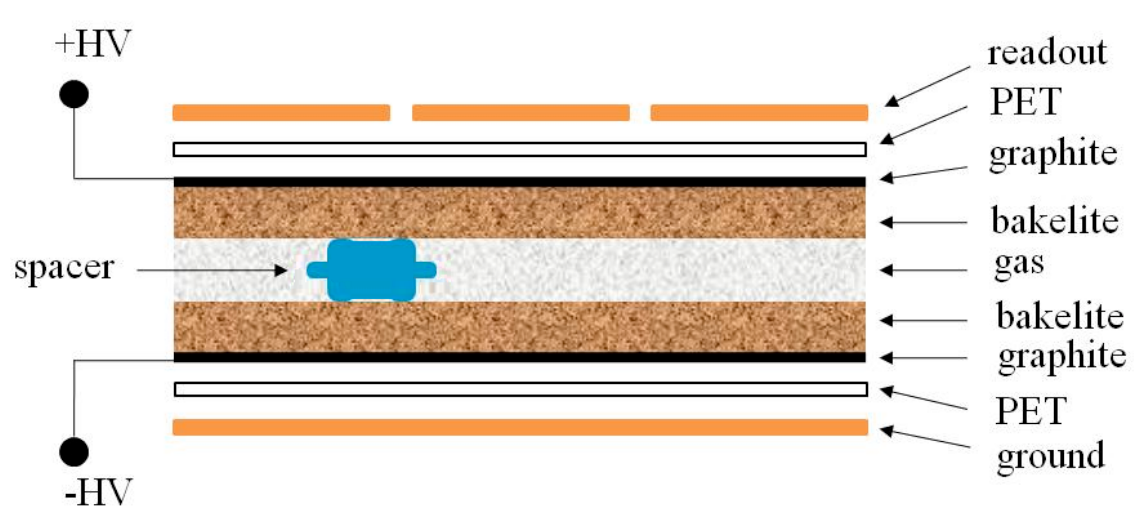}
    \caption{
      The basic structure of a bare RPC.
      The button spacers maintain a uniform gap, but create small dead areas.
      The PET layers are a 100~$\mu$m polyethylene terephthalate film covering the graphite.
    }
    \label{Fig4}
  \end{center}
\end{figure}

Resistive plate chambers (RPCs) are gaseous particle detectors that consist of two resistive planar electrodes
separated by a gas gap~\cite{RPCbasic}.
The Daya Bay RPCs are similar to those used by BESIII~\cite{BES}.
The RPC electrodes are 2~m $\times$ 1~m Bakelite sheets with Melamine surfaces, but,
unlike similar RPCs, linseed oil is not applied to the inner surface~\cite{no_oil}.
The outside of the RPC electrodes is coated with graphite for uniform HV distribution.
The gas gap between the two electrodes is 2.0 mm.
Daya Bay operates the RPCs in streamer mode, which provides relatively large signals and thereby simplifies the readout electronics.
Figure \ref{Fig4} is a schematic of a bare RPC,
the functional element of the Daya Bay RPC modules.

\begin{figure}
  \begin{center}
    \includegraphics[width=7.8cm]{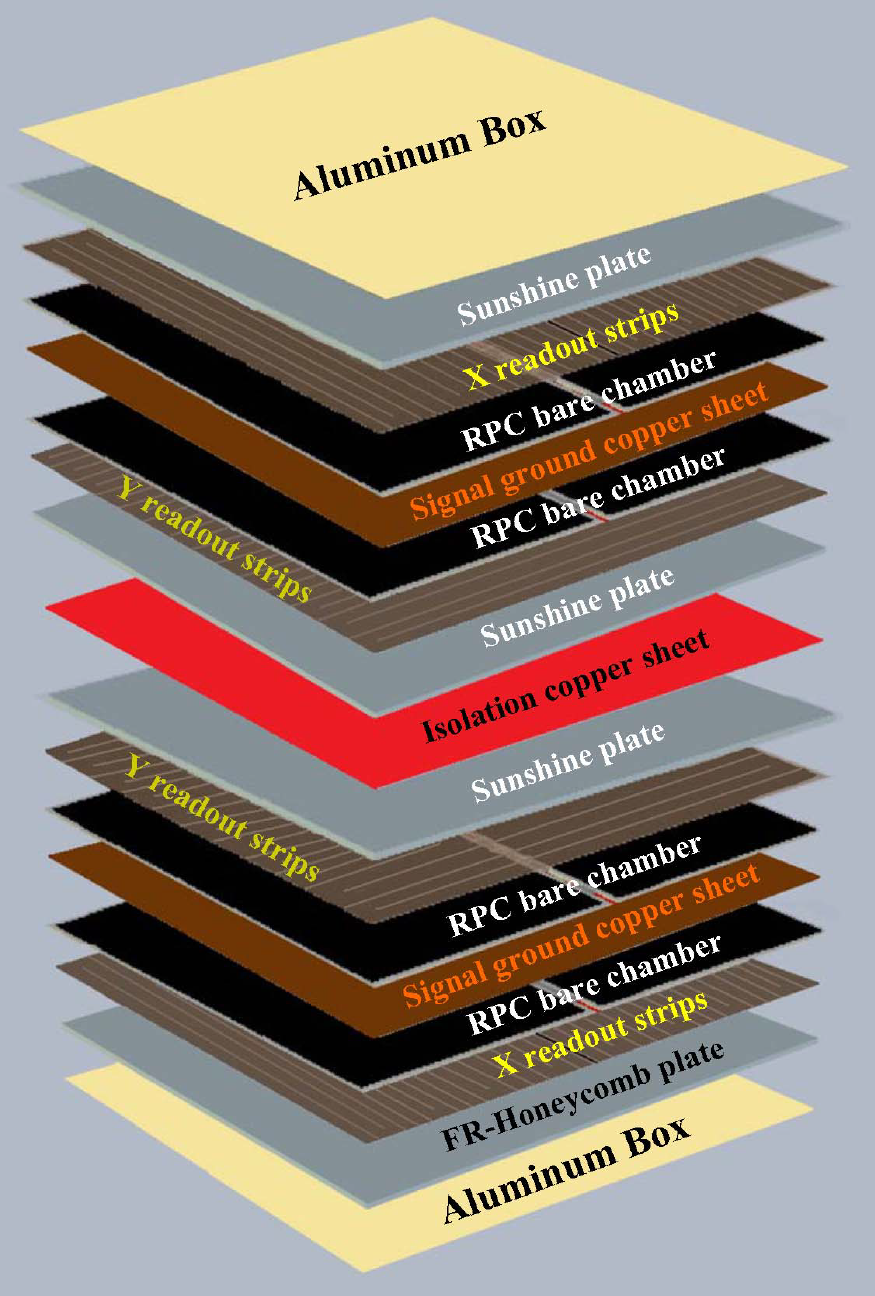}
    \caption{ A schematic of the RPC module structure.
              Each module has four pairs of side-by-side bare RPCs.
              The ``Sunshine'' plates are 1.0~cm twin-wall polycarbonate.
    }
    \label{Fig5}
  \end{center}
\end{figure}

\begin{figure}
  \begin{center}
    \includegraphics[width=7.8cm]{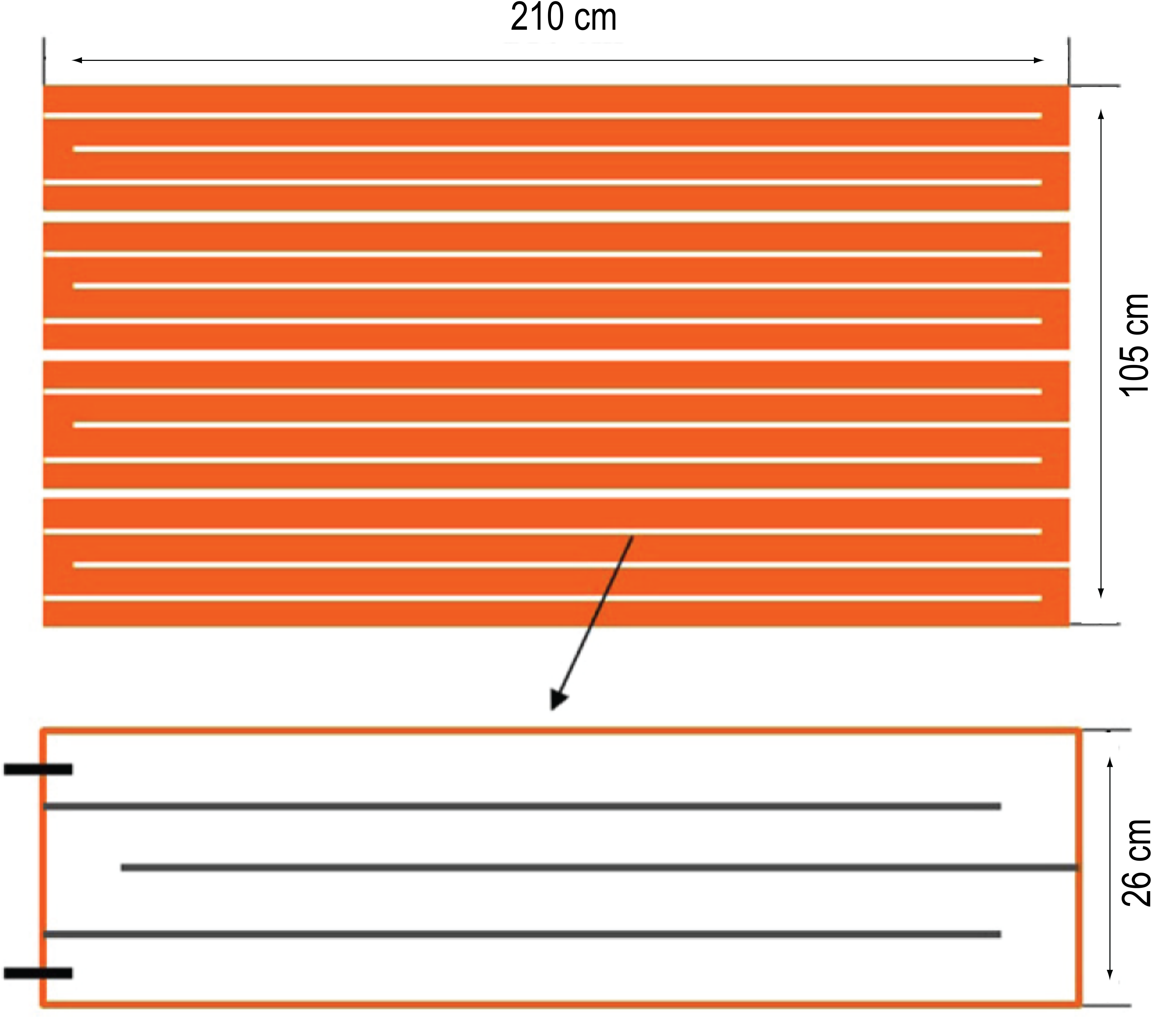}
    \caption{
      Top: The RPC zigzag (or folded) readout pattern, showing four of eight readout strips.
      Bottom: A single readout strip, vertically expanded to improve visualization of the folds,
      and showing the readout contacts on the left.
      The separation between readout strips is 0.25~cm.
      The folds in a single readout strip are 6.25~cm wide and separated by 0.25~cm.
    }
    \label{Fig6}
  \end{center}
\end{figure}

RPC modules, 2.17~m $\times$ 2.20~m $\times$ 8~cm,
are constructed from four pairs of side-by-side bare RPCs arranged in layers (Fig.~\ref{Fig5}),
each separated by insulating materials, support panels, ground planes,
and copper-clad FR-4 readout planes associated with each pair of side-by-side bare RPCs,
inside an aluminum box.
The readout planes each have eight 26~cm $\times$ 2.10~m readout strips oriented like $X~Y~Y~X$.
The strips have a zigzag design (Fig.~\ref{Fig6}), equivalent to strips 6.25~cm wide and 8.4~m long.

Each hall has an RPC gas system 
which mixes argon, freon (R134a), isobutane, and SF${}_6$
in the ratio 65.5:30.0:4.00:0.500~\cite{rpc_gas_mixture}.
Each gas system comprises a mixing panel, a main gas distribution panel,
a fire/gas safety monitoring system,
an MKS Instruments 247D mass flow control system,
a Varian GC430 gas chromatograph,
and gas supplies in a dedicated utility room.
The gas from the main mixing panel is distributed by two independent gas channels
to the upper and lower halves of each module,
with groups of four bare RPCs connected in series.
If one gas channel fails, the other will continue to supply one $X$ and one $Y$ readout.
The exhaust of each gas channel is fed into mineral-oil bubblers in racks mounted on the RPC support frame,
where the bubbles are electronically counted and recorded.
The nominal gas flow rate is about 1 volume per day.
The gas mixture is analyzed every two hours by the chromatograph, ensuring the correct gas mixture.
The high ambient humidity at Daya Bay, greater than 65\%, is absorbed through the RPCs,
thereby making the gas mixture sufficiently humid without adding water vapor,
which the gas system is capable of doing.

The RPC HV system consists of a CAEN 1527LC mainframe in each hall's dedicated electronics room,
populated with equal numbers of positive (+6~kV A1732P) and negative (-4~kV A1733N) 12-channel HV cards,
with fanout boxes and HV interface boxes in the experimental hall.
The systems in EH1 and EH2 each use four CAEN HV cards, while there are six in EH3.
Each HV channel is divided into 9 channels by a fanout box and distributed to RPCs via 
RG59 cables through an HV interface box mounted on each RPC module.
Each RPC layer of a module is connected to one positive and one negative HV channel.
If one of the high voltage channels in a module fails, three of the four layers will continue to function.

Between January 2008 and July 2009, Gaonengkedi Ltd. Co. (Beijing) produced
756 2.1~m $\times$ 1.1~m and 756 2.1~m $\times$ 1.0~m bare RPCs for the main RPC arrays in the three experimental halls,
a total area of about 3300~m$^2$.
An additional 24 of each size of bare RPCs were produced for six special modules used to form small RPC telescopes.
Concurrent with the production,
the bare RPCs were tested using cosmic muons at IHEP,
where the average efficiency was 96.1\%, the noise rate was 0.15 Hz/cm$^2$, and
the dark current was 2.5 $\mu$A/m$^2$~\cite{quality}.
The characteristics of these RPCs are in agreement with previous experience~\cite{rpc_eff}.

The bare RPCs that passed all tests
(about 12\% were rejected) were then assembled into RPC modules and performance-tested at IHEP~\cite{rpc_study}.
The intersection area of two orthogonal $X$ and $Y$ readout strips defines a 26 cm $\times$ 26 cm patch.
The average efficiency of each patch is about 99.8\% with a coincidence between any two of the four layers (2/4).
The average patch efficiency is about 98.0\% with a coincidence between any three of the four layers (3/4).
Considering the dead areas created by the button spacers (Fig.~\ref{Fig4})
and bare RPC frames running through the center of each layer (Fig.~\ref{Fig5}),
this agrees well with the simulated module 3/4 efficiency based on the measured average bare RPC efficiency of 96.1\%.
The accepted modules were transported in crates to Daya Bay by truck in several trips during the dry months of 2010 and 2011.
Prior to shipping, a determination was made that the effects of vibrations during transport would be negligible.

\subsection{Construction and Installation}
\label{subCandI}

\begin{figure}
  \begin{center}
    \includegraphics[width=7.8cm]{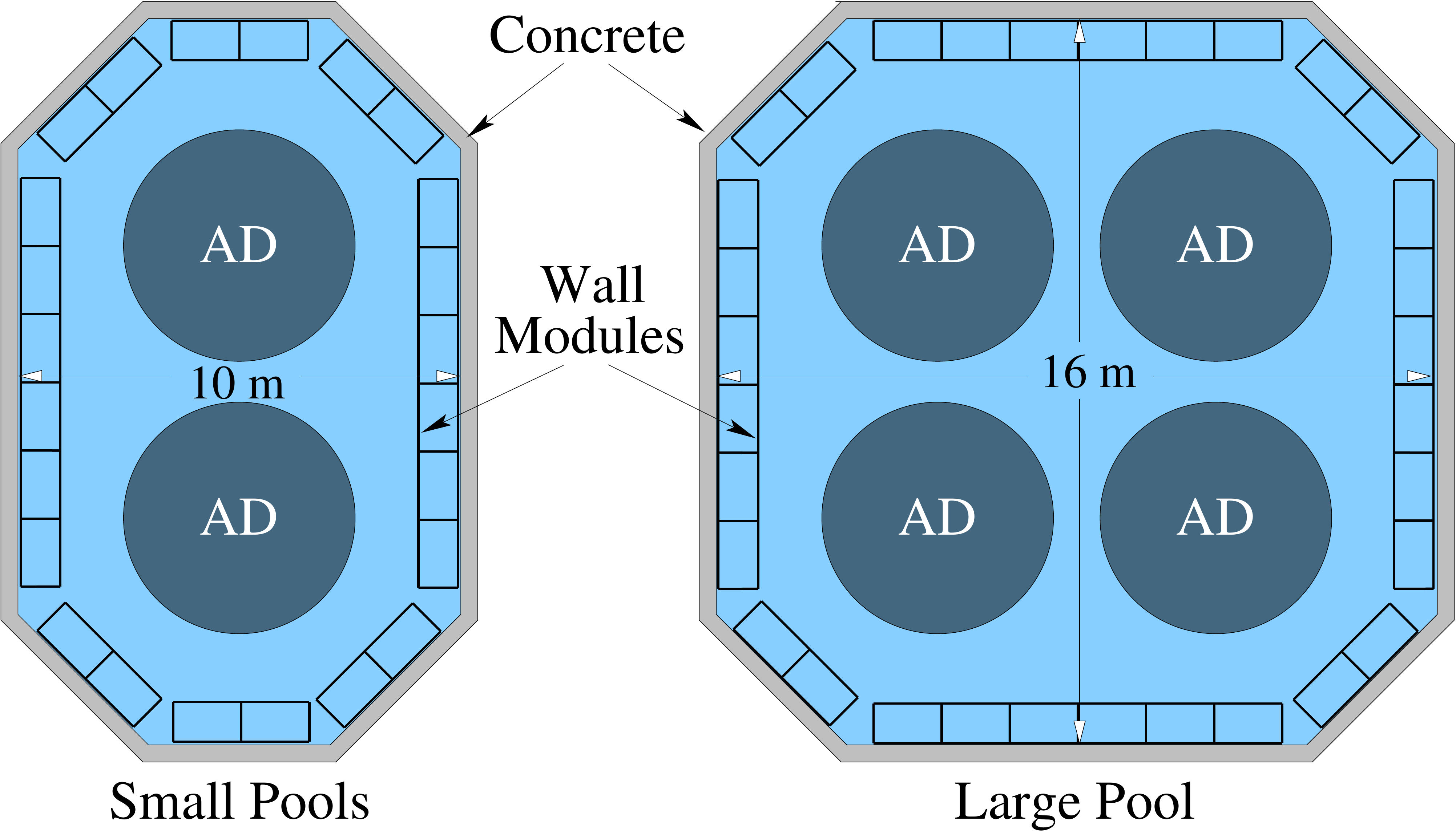} 
    \caption{
      Sketch of the small (near halls) and large (far hall) pools, all 10~m deep.
      The wall modules supporting the PMTs are 1~m wide (Fig. \ref{Fig8}).
      The floor framework under the ADs is not shown.
      The OWS consists of the volume contained within the wall modules and
      includes the bottom 1~m just above the pool floor.
      The IWS includes everything else, and extends to the surface of the pool.
      An optical barrier (not shown) covering the wall modules and floor framework isolates the IWS and OWS from each other.
    }
    \label{Fig7}
  \end{center}
\end{figure}

The pools are octagonal, as shown in Fig. \ref{Fig7},
and constructed of poured concrete reinforced with grounded rebar.
Each pool has a small sump pit about $1$~m${}^2$ by $1$~m deep near one corner,
which is used to house a turbine pump for draining.
The pool walls extend about 20~cm above the experimental hall floor,
forming a concrete curb about 20~cm wide around the pool.
The two corner walls nearest the utility rooms in each hall have several 10 and 15~cm
penetrations consisting of PVC pipes embedded in the concrete and above the water level,
through which pass all of the cables, plumbing, and
lines from water level and temperature sensors.

Following a survey of the pool, pairs of anchor holes were drilled
into the concrete pool walls at marked locations on a roughly 2~m
$\times$ 2~m grid, into which steel anchors were epoxied.
Each such pair of anchors supports a type-304 stainless steel anchor pad.
The anchor pads on each of the eight walls were adjusted with leveling
bolts to be co-planar with each other, then grouted in place. 
The entire concrete surface of the pool (walls and floor)
was then pressure-washed to remove the laitance which loosely covers newly
poured concrete.
When the concrete was sufficiently dry (less than 4\% moisture), three coats of \PermaFlex
urethane-based paint were applied.
As with all materials used in the pools in Daya Bay, the PermaFlex
was extensively tested for compatibility with ultrapure water.
Besides being a durable, waterproof coating for the concrete, it
is also an effective barrier against radon penetration. 
To ensure adequate and uniform application, each successive coat was a different color.
The color of the top coat plays no role in the optical properties of the pools because all surfaces are
covered with Tyvek.
In addition to being compatible with ultrapure water and nearly opaque,
Tyvek is highly reflective~\cite{Tyvek01,Tyvek02,Tyvek03}, which increases the pool light collection efficiency
albeit at some cost to signal timing.
Daya Bay uses a highly reflective multi-layer film formed from two pieces of 1082D Tyvek ~bonded with a layer of polyethylene,
for which the reflectivity in air is more than 96\% for wavelengths from about 300 to 800~nm.
The reflectivity is 99\% in water~\cite{IHEPprototype}.
The reflectance is diffuse with a small specular component.

\begin{figure}
\begin{center}
  \includegraphics[width=7.8cm]{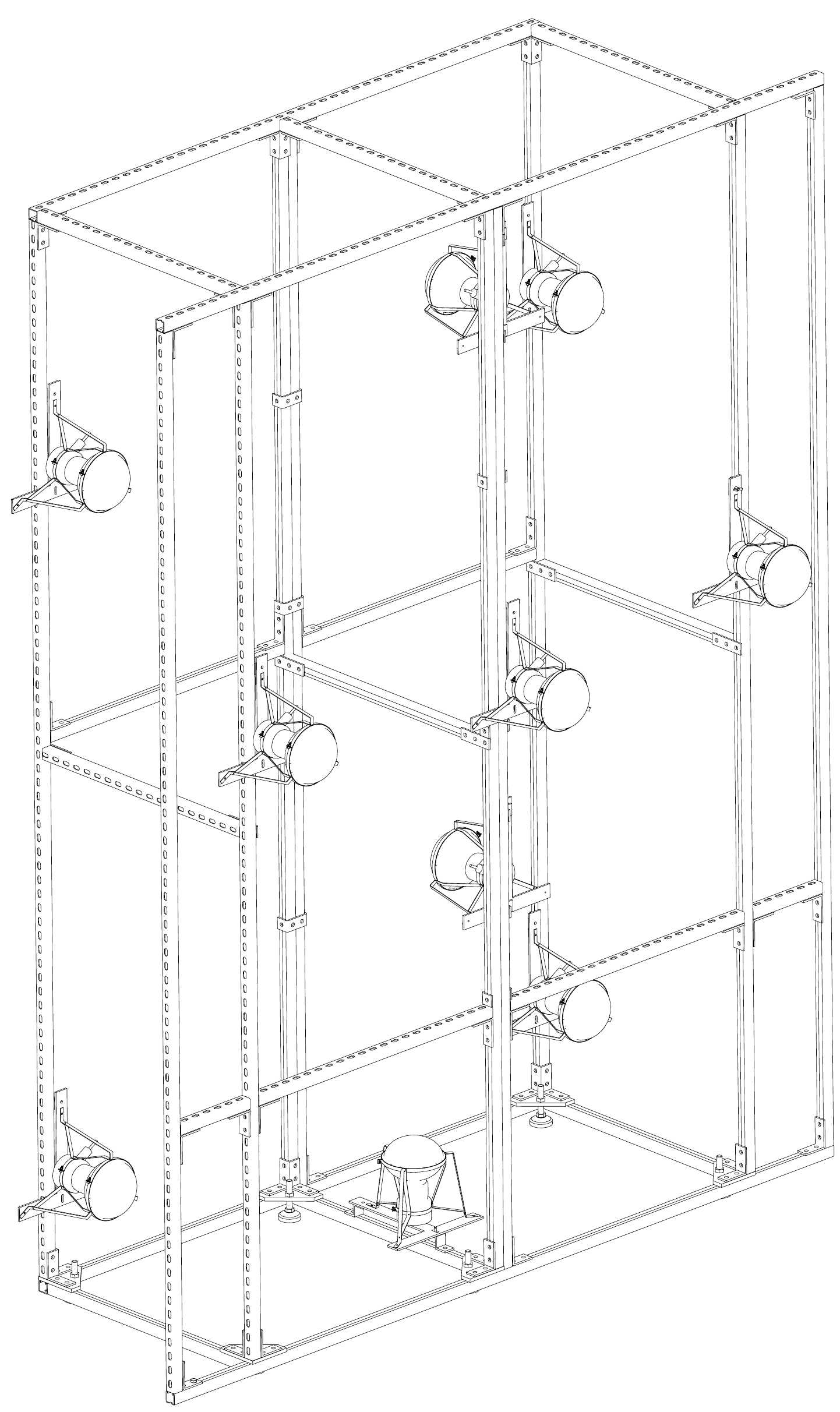}
  \caption{
    Typical wall module.
    Shown is the bottom module for one of the short walls in the small pools (top and bottom of Fig. \ref{Fig7}, left).
    The ``outer edge'' (the back, facing away and to the left) is fastened to the wall anchor pads.
    The bottom modules are all 4~m high and are leveled with adjustable feet which rest on the pool floor.
    The middle and top modules are all 3~m high, and rest on, and are fastened to, the modules below them.
    Each PMT (``Tee'') assembly is supported by a pair of L-shaped pieces (not clearly visible)
    mounted on the Unistrut.
    The three PMTs on the ``inner edge'' (the front, facing forward and to the right), just above middle-height, are in the IWS.
    The other PMTs are in the OWS.
    The two OWS PMTs mounted on the vertical center support, facing away and to the left, face outward (i.e., away from the ADs),
    while the other OWS PMTs face inward (see Table \ref{Tab1}).
  }
  \label{Fig8}
\end{center}
\end{figure}

After the PermaFlex had cured, Tyvek was draped down over the pool walls in 2~m wide strips
and extended out over the pool floor.
Each strip had an extra 12~m at the top of the pool which was folded, covered, and stored on the hall floor
for use in a later installation step.
The seams between the strips were all heat-welded. 
Pre-assembled wall modules (Fig. \ref{Fig8})
were then lowered into place and fastened to the anchor pads on the pool walls.
Each section of the wall framework comprises three stacked, pre-assembled modules
made from type-304 stainless steel Unistrut~\footnote{Reg. trademark. See, e.g., http://www.unistrut.com.}.
All of the materials in the modules, including the fittings, PMT brackets and ``Tees'', and other hardware,
are type-304 stainless steel, and were treated with acid in a process known as ``pickling'', which
is necessary to restore stainless-steel surfaces to their full, corrosion-resistant state following any kind of heating,
bending, or machining (any of these actions render stainless steel susceptible to corrosion, especially in ultrapure water).

PMT assemblies were then installed on the wall modules
in top-down order to minimize risk to installed PMTs from the possibility of damage from falling objects during installation.
Because some of the EMI PMTs imploded during pressure testing, 
indicating that they were less able to withstand pressure than the Hamamatsu PMTs,
they were deployed only in the top 2~m of the pool.

Once the OWS PMTs were all installed on the wall modules,
the Tyvek strips' extra 12~m which had been stored at the top, outside the pool,
was fixed to the outer edge of the wall modules and draped down
over the inner edge of the wall modules, extending down to one meter above the pool floor~\footnote{
  Recall that the OWS surrounds the IWS on the sides and bottom, and is everywhere 1~m wide.}.
As before, the seams between the strips were heat-welded.
This optically isolates the IWS and OWS at the walls.
Cable trays were then mounted on top of the framework, over the Tyvek.
Small perforations were made in the Tyvek for fastening the cable tray and IWS PMTs to the Unistrut.

The IWS PMTs were then installed on the wall modules.
After a pause during which the AD stands were installed, Unistrut was assembled in-place to form the floor framework.
Like the wall modules, the floor frames were also leveled with adjustable feet.
Following the installation of the OWS floor PMTs,
the floor framework was covered with Tyvek, completing the optical isolation of the IWS and OWS,
then the IWS floor PMTs were installed.
At this point, the ADs were installed on their stands,
and the penetrations in the curb were sealed with a waterproof, opaque sealant.

After a brief commissioning dry run, the pools were filled with water
and covered with a 0.5 mm-thick gas- and light-tight black rubberized cloth supported on stainless steel cables.
With the cover and penetration sealing in place, the pools are light-tight and sufficiently gas-tight so that a
positive-pressure cover gas of dry nitrogen prevents exposure of the water to air.
The flow rate is about 80~L/min for the small halls, and 100~L/min for the large.

In parallel with the installation activities described above,
HV interfaces and Front End Cards (FECs) were installed on the RPC modules on site,
which were then tested for gas tightness and HV integrity before installation. 
The RPCs were required to hold 20~cm of water overpressure with a drop rate of less than 2\% per day.
8~kV was applied to the RPC modules while they were flushed with pure argon to check the HV connections and basic performance.
Readout strip and ground connections were also checked.
The RPC modules were then transported to the experimental halls in the same containers in which they were shipped from Beijing.

\begin{figure}
\begin{center}
    \includegraphics[width=7.8cm]{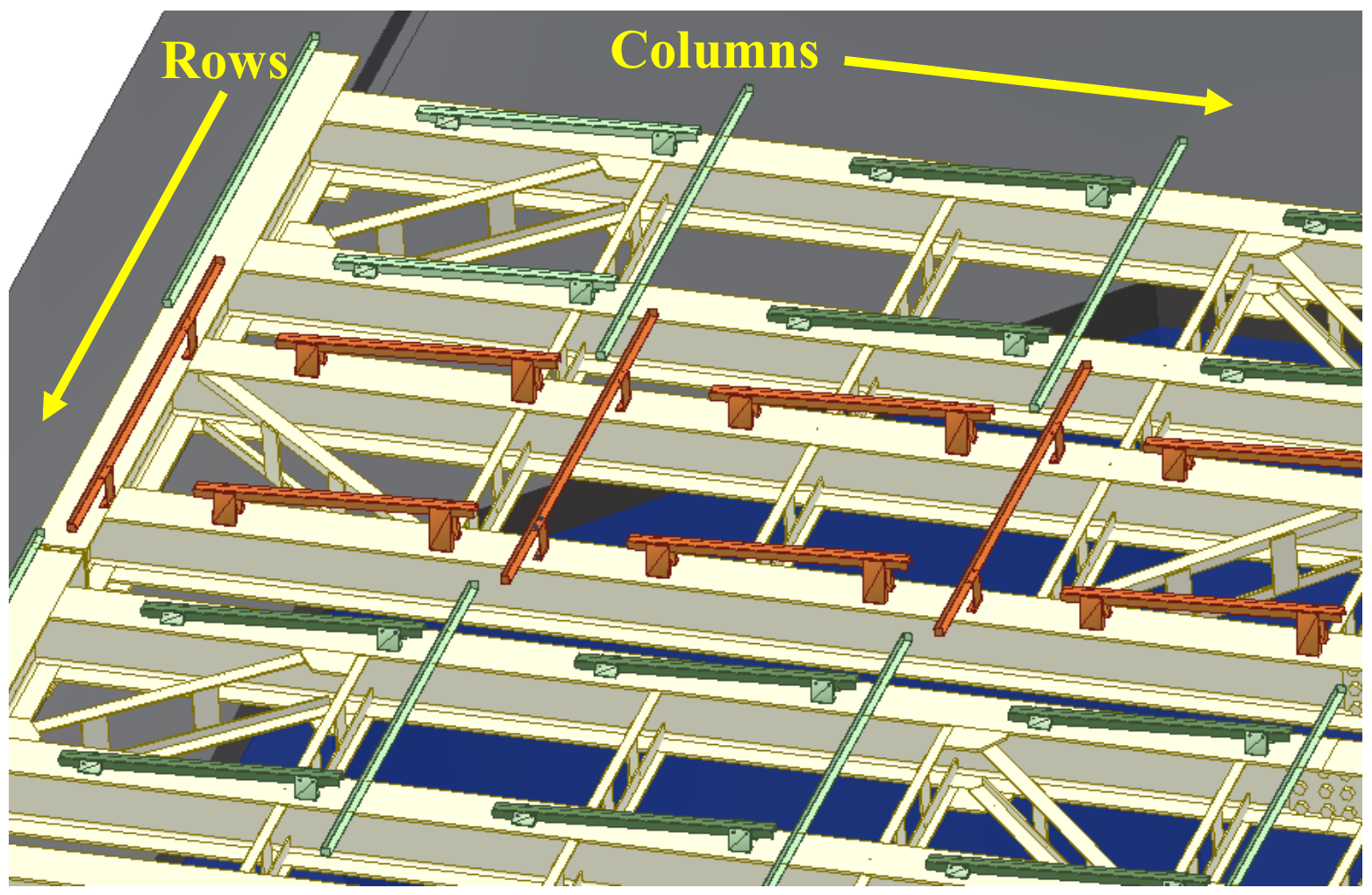}
    \caption{
      The RPC modules are installed on a support structure in two parallel, partially overlapping layers.
      Modules are mounted on pairs of the brown supports (higher layer)
      or on pairs of the green supports (lower layer).
      Modules in a row alternate between the higher and
      lower supports, providing overlap in the row-direction.
      The supports are canted at 3${}^\circ$, providing overlap in the column-direction.
    }
    \label{Fig9}
\end{center}
\end{figure}

The RPC modules were laid on a steel support structure in a staggered pattern such that there is
10 cm overlap between neighboring modules to minimize dead regions (Fig. \ref{Fig9}).
The RPC support structure is installed on rails,
so that the RPCs can be rolled away to allow access to the pool.
There are 6 $\times$ 9 modules in EH1 and EH2, and 9 $\times$ 9 modules in EH3.  
 
Two RPC modules were specially installed in each experimental hall to form the RPC telescopes.
These modules are about 2 m above the RPC array, at the middle on opposing sides of the pool, and partially overlap the RPC array.
Muons that pass through both the telescope and the main RPC array can be tracked with good angular resolution,
as described in \S~\ref{SubMuonDist}.

\subsection{Readout and Triggering}
\label{SubReadout}
All hardware related to the powering or readout of detector signals is connected to a Signal Ground that includes
the cable trays, wall modules,
electrodes in the rock, a copper grid in the concrete floor outside the pool, and the rebar of the pool.
A separate Safety Ground is brought into the halls with
the AC power, connected to various peripheral systems such as the 
Detector Controls System~\cite{DCS} and networking. 

The Muon System PMTs' signals are fed into ADCs and TDCs.
These, and their readout, are the same as those of the AD PMTs~\cite{ADpaper,LocalTriggerBoard,DAQarch}.
An IWS or OWS PMT readout (the two are independent) is initiated by one or more of the following triggers:
\begin{itemize}
\item[$\bullet$] Multiplicity Trigger: The PMT multiplicity (the number of PMTs each with charge above
a preset threshold) meets or exceeds a preset minimum.
Only PMTs above 0.25~pe are read out.
\item[$\bullet$] Energy Sum Trigger: The total PMT charge exceeds a preset total-charge threshold.
Only PMTs above 0.25~pe are read out.
\item[$\bullet$] Periodic Trigger: Unbiased at 10$\,$Hz.
All PMTs are read out.
\end{itemize}
\smallskip

The 32 signal strips in each RPC module are read out and discriminated by one FEC.
Up to 15 FECs are read out by one Read Out Transceiver located on the RPC support frame~\cite{rpc_readout}.
RPC module triggers are collected and communicated to the data acquisition system by
an RPC Trigger Module and Read Out Module in the electronics room~\cite{rpc_VME}.
An RPC readout is initiated by one of the following triggers:
\begin{itemize}
\item[$\bullet$] 2/4 Trigger: Any two of the four layers in a module sees a signal above threshold.
Only the triggered modules are read out.
\item[$\bullet$] 3/4 Trigger: Any three of the four layers in a module sees a signal above threshold.
Only the triggered modules are read out.
\item[$\bullet$] Periodic Trigger: Unbiased at 10$\,$Hz.
All modules are read out.
\end{itemize}
\smallskip

The Muon System performs its function as a veto strictly in the offline environment.
The presence or absence of an IWS, OWS, or RPC trigger does not affect the readout of the AD PMTs.
AD readouts are matched with IWS, OWS or RPC readouts with GPS-derived time stamps,
accurate to 25~ns.
For the IBD event selection,
an event is tagged as a muon if either the calibrated total energy in an AD is greater than $20$~MeV
or if the PMT multiplicity in the IWS or OWS is greater than 12 (considerably higher than the multiplicity trigger requires).
The RPCs are not used to tag muons in the oscillation analysis,
though they are used in cosmogenic background studies.

\subsection{Monitoring and Control}
The Daya Bay Detector Controls System~\cite{DCS} provides the means of remotely monitoring and controlling the detectors'
operating parameters for each hall.
The water system monitors include water temperature, oxygen content, water level and resistivity.
The HV of each water pool PMT is remotely controlled and monitored, and the currents monitored.

The RPC gas system monitors include the flow rates of the
gas mixture and each of its components, gas tank pressures and weights, gas feed and return humidity, and several status
conditions, such as the state of the hazardous gas monitor (which would detect an isobutane leak in the gas room)
and interlock status.
If the flow rate of any gas component is not within its specified range, the gas system will shut down automatically and 
send an alarm signal to monitoring personnel.
The gas system can also be shut down remotely.

The HV of each RPC channel is remotely controlled and monitored, and the dark currents are monitored.
The HV and gas systems are interlocked through the remote monitoring program so that, in the case of an alarm signal 
from the gas system, the HV monitoring program will warn monitoring personnel and allow them 30 minutes to resolve the 
issue before automatically turning off the RPC HV.

The Daya Bay Performance Quality Monitoring system~\cite{ODM} provides the means of monitoring the performance of the detectors,
continuously displaying the detector trigger rates, individual TDCs and ADCs, and detector channel-maps, to name a few.

\subsection{Commissioning}

After the PMTs were all installed in the EH1 pool,
and before installing the ADs and filling with water,
a dry run was conducted to perform
a full test of everything from HV, PMT, readout electronics, and the data acquisition system.
A cover was placed over the pool, and HV gradually applied to the water pool PMTs while checking for light leaks.
As expected because of their recent exposure to bright light, the PMT dark rates were initially elevated.
After a few days the PMTs quieted down somewhat, but their rates were still substantially higher than expected.
One contribution to the elevated rates
was unanticipated fluorescence in the insulating material (polyethylene) on the AD stands,
which scintillates when struck by $\alpha$ particles from radon decay.
Once this problem was identified,
the AD stands in EH1 and EH2 were covered with opaque material, which reduced the PMT rates by about 35\%,
while a different insulating material was used in EH3, where the AD stands had not yet been installed.
With this covering in place,
the PMT rates (threshold 0.25~pe) were typically 4 to 5~kHz in the OWS, and 6 to 8~kHz in the IWS.
This is to be compared to what had been measured in the controlled environment of the lab,
where the typical rates were about 1~kHz.
This is almost certainly due to 
$\alpha$ particles from radon decay, which ionize and excite atmospheric nitrogen~\cite{RnAir1},
producing light
in the PMT-sensitive range of 300-500~nm~\cite{RnAir2,RnAir3}.
Given that the volume of air in the IWS is so much larger than that in the OWS,
it is not surprising that the IWS rates were substantially higher than the OWS rates in the dry pool.
At the end of the dry run HV was turned off, the cover removed, and the ADs installed.

Once each pool was filled with water and covered,
the water pool PMTs' HVs were gradually raised to their full nominal settings while testing for light leaks.
For reasons not entirely understood, a PMT's dark rate generally increases when it is immersed in water,
even in ultrapure water.
Although the PMTs were required to have dark rates less than $10$~kHz in the laboratory,
some dark rates remained at $15$~kHz after the pools were filled and even after a two-week period of darkness.
The typical rates in EH1 were about 9~kHz for the IWS and about 14~kHz for the OWS.
This difference is expected since the OWS PMTs are closer to the granite of the pool walls
which emit quite high rates of gamma radiation.
The water in the OWS shields the IWS from most of these gammas.
This effect is negligible when the pool is dry, because it operates by generating Cherenkov light in the water.
Based on initial data taken over the first month, the IWS and OWS thresholds for the multiplicity trigger
and charge thresholds for the energy sum trigger were set as shown in Table \ref{tab:pmtThresh}.
\begin{table}
  \centering
  \caption{
    IWS and OWS trigger thresholds, as of October 22, 2012. 
    For the multiplicity trigger, the single PMT charge threshold is 1.2~mV,
    which corresponds to 0.25~photoelectrons (pe).
  }
  \label{tab:pmtThresh}
  \vspace{5pt}
  \begin{tabular*}{7cm}{c@{\extracolsep\fill}cc}
    \hline
    \hline
    \hskip 0.5cm {\sf Pool}\hskip 0.5cm\phantom{.}
             & {\sf Multiplicity}\hskip 0.5cm\phantom{.} & {\sf Energy Sum}\hskip 0.2cm\phantom{.}  \vspace{-5pt} \\
             & {\sf Threshold ($\ge$)}\hskip 0.2cm\phantom{.} & {\sf Threshold}\hskip 0.2cm\phantom{.}  \\ \hline
    EH1 $\begin{matrix}\mbox{IWS}\\ \mbox{OWS}\end{matrix}$
             & $\begin{matrix}\mbox{6}\hskip 0.3cm\phantom{.}\\
                              \mbox{7}\hskip 0.3cm\phantom{.}\end{matrix}$
                                       & $\begin{matrix}\mbox{ 8.9~mV (1.8~pe)}\hskip 0.3cm\phantom{.}\\
                                                        \mbox{10.0~mV (2.0~pe)}\hskip 0.3cm\phantom{.}\end{matrix}$\\ \hline
    EH2 $\begin{matrix}\mbox{IWS}\\ \mbox{OWS}\end{matrix}$
             & $\begin{matrix}\mbox{6}\hskip 0.3cm\phantom{.}\\
                              \mbox{7}\hskip 0.3cm\phantom{.}\end{matrix}$
                                       & $\begin{matrix}\mbox{ 8.9~mV (1.8~pe)}\hskip 0.3cm\phantom{.}\\
                                                        \mbox{10.0~mV (2.0~pe)}\hskip 0.3cm\phantom{.}\end{matrix}$\\ \hline
    EH3 $\begin{matrix}\mbox{IWS}\\ \mbox{OWS}\end{matrix}$
             & $\begin{matrix}\mbox{6}\hskip 0.3cm\phantom{.}\\
                              \mbox{8}\hskip 0.3cm\phantom{.}\end{matrix}$
                                       & $\begin{matrix}\mbox{12.2~mV (2.4~pe)}\hskip 0.3cm\phantom{.}\\
                                                        \mbox{14.4~mV (2.9~pe)}\hskip 0.3cm\phantom{.}\end{matrix}$\\
    \hline
    \hline
  \end{tabular*}
  \centering
\end{table}

Because the underground temperature and humidity were considerably higher than at IHEP,
the RPC dark current at Daya Bay was initially too high at the HV value of 7.6~kV used in the tests at IHEP.
To reduce the ambient humidity around the RPCs, and thus the dark current, each hall was retrofitted with 
a dry air system that blows dry air into the RPC modules.
This immediately reduced the dark current by a factor of two.
After steady long-term operation with the new dry air system,
the additional reductions in singles rates and dark current eventually allowed
the RPCs to be operated at their nominal HV value of 7.6~kV. 
Based on threshold scans in each hall, the signal thresholds were all set to 35~mV.
The threshold was chosen to balance efficiency and noise.
The threshold was selected to be higher than the IHEP test setting of 30~mV to reduce the impact of accidentals and other noise. 
It was determined that the 2/4 trigger produced an excessive volume of RPC data,
so the RPCs are read out with only the 3/4 trigger and periodic trigger during Physics data taking,
with a loss in efficiency on the order of 1\%.

\section{Muon and detector simulation}
\begin{figure}
\begin{center}
    \includegraphics[width=7.8cm]{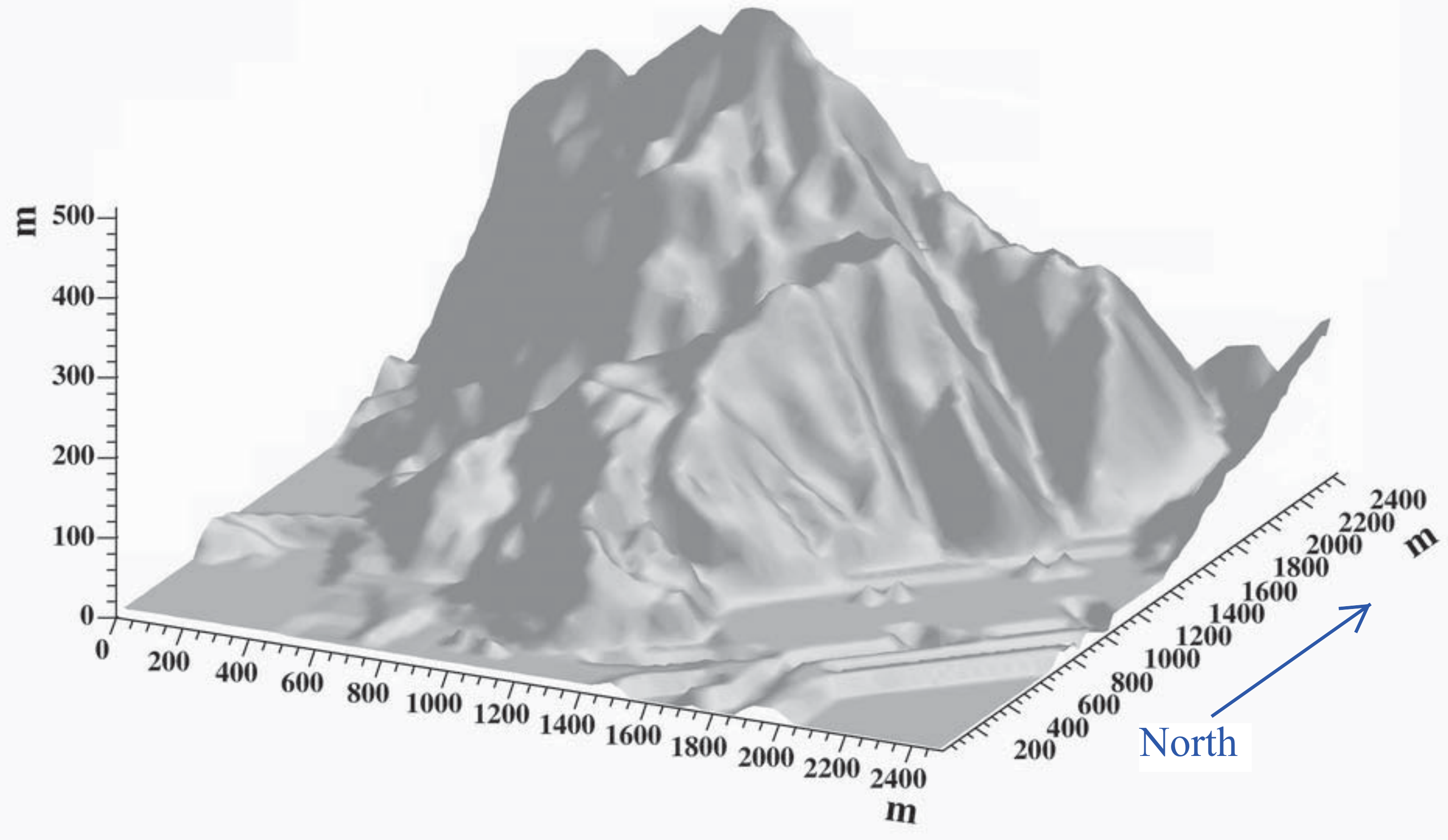}
    \caption{
      3D image of the Daya Bay area generated by ROOT~\cite{ROOTarticle} from the digitized mountain profile.
      The tunnel entrance is at the base of the mountain at the left side of the figure.
    }
    \label{Fig10}
\end{center}
\end{figure}

The muon flux at sea-level is reasonably well-described by Gaisser's formula~\cite{Gaisser:1990vg,gaisser2},
originally introduced in 1990.
However, for our purposes it is necessary to modify this formula~\cite{modified}
to better describe the low energy spectrum,
to account for the earth's curvature,
and to better describe large-zenith muons~\cite{Chirkin:2004ic,Volkova,Kellogg:1978gn,Muraki:1983sj}.
High and low precision topographic maps of the region of the Daya Bay Nuclear 
Power Plant were merged to obtain a realistic, digitized mountain profile (Fig.~\ref{Fig10}).

\begin{figure}
\begin{center}
    \includegraphics[width=7.8cm]{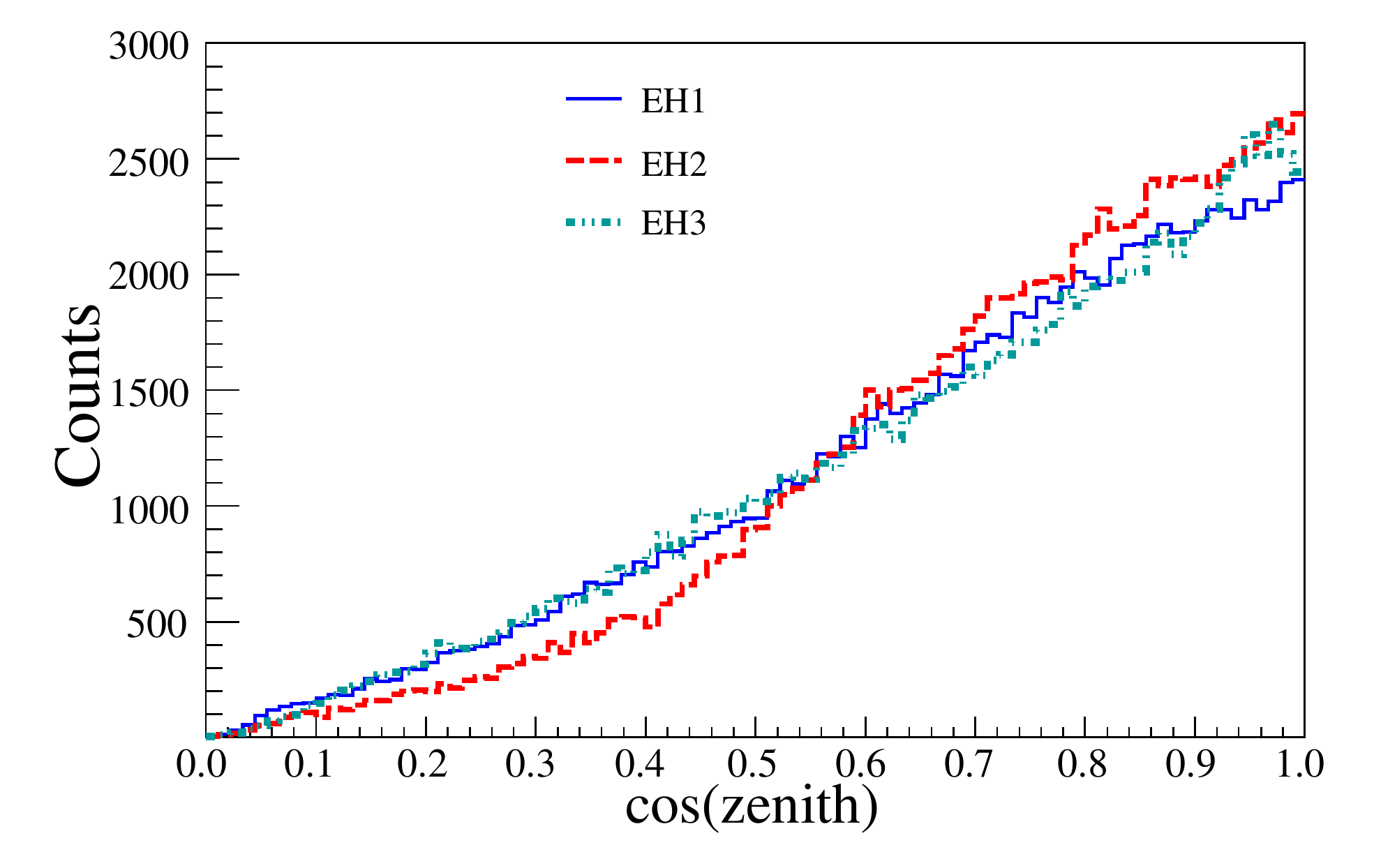}
    \includegraphics[width=7.8cm]{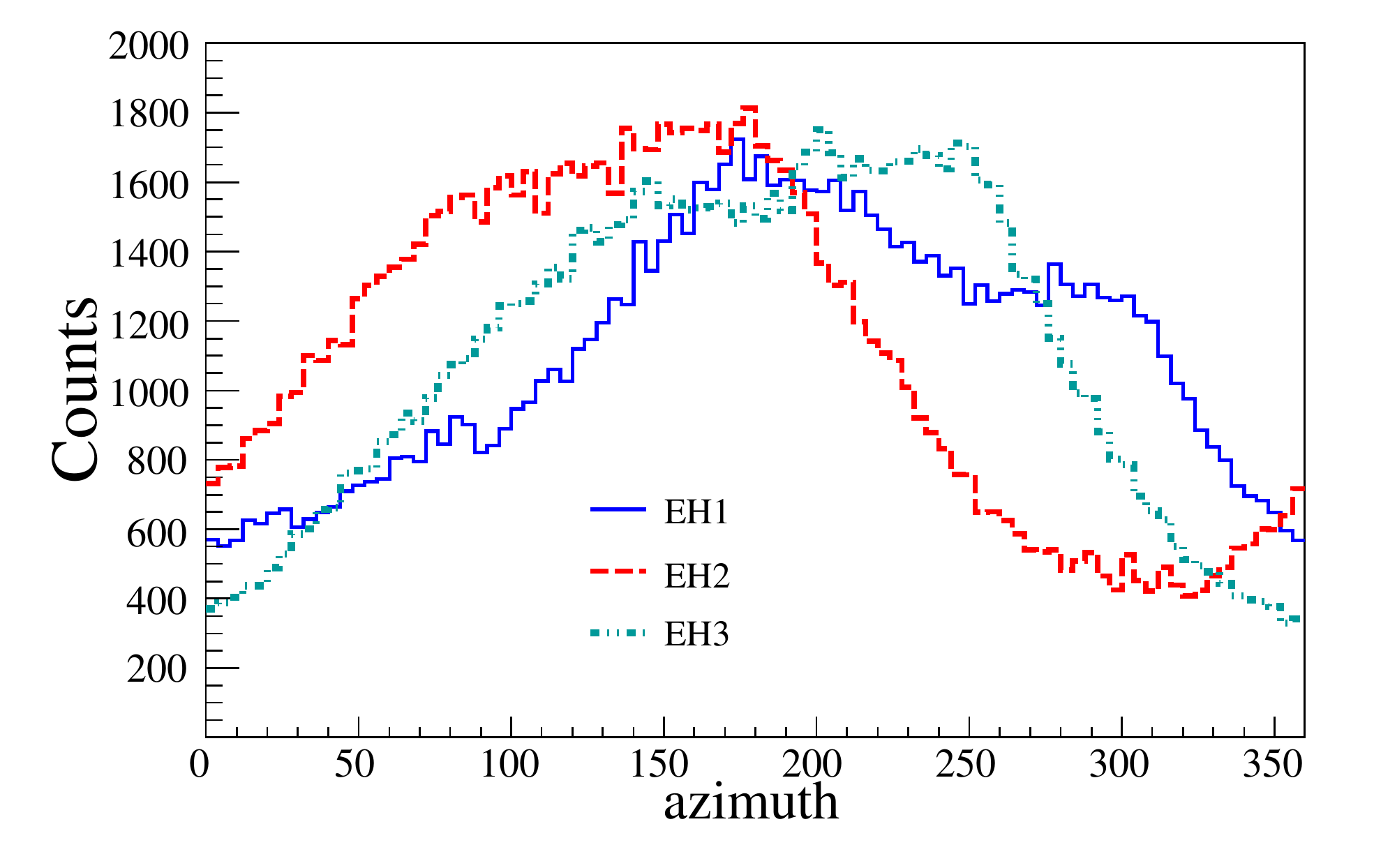}
    \caption{
      Simulated muon trajectories.
      By definition, zenith is the angle from vertical and
      azimuth is the horizontal compass angle from true North.
    }
    \label{Fig11}
\end{center}
\end{figure}

\begin{figure}
\begin{center}
    \includegraphics[width=7.8cm]{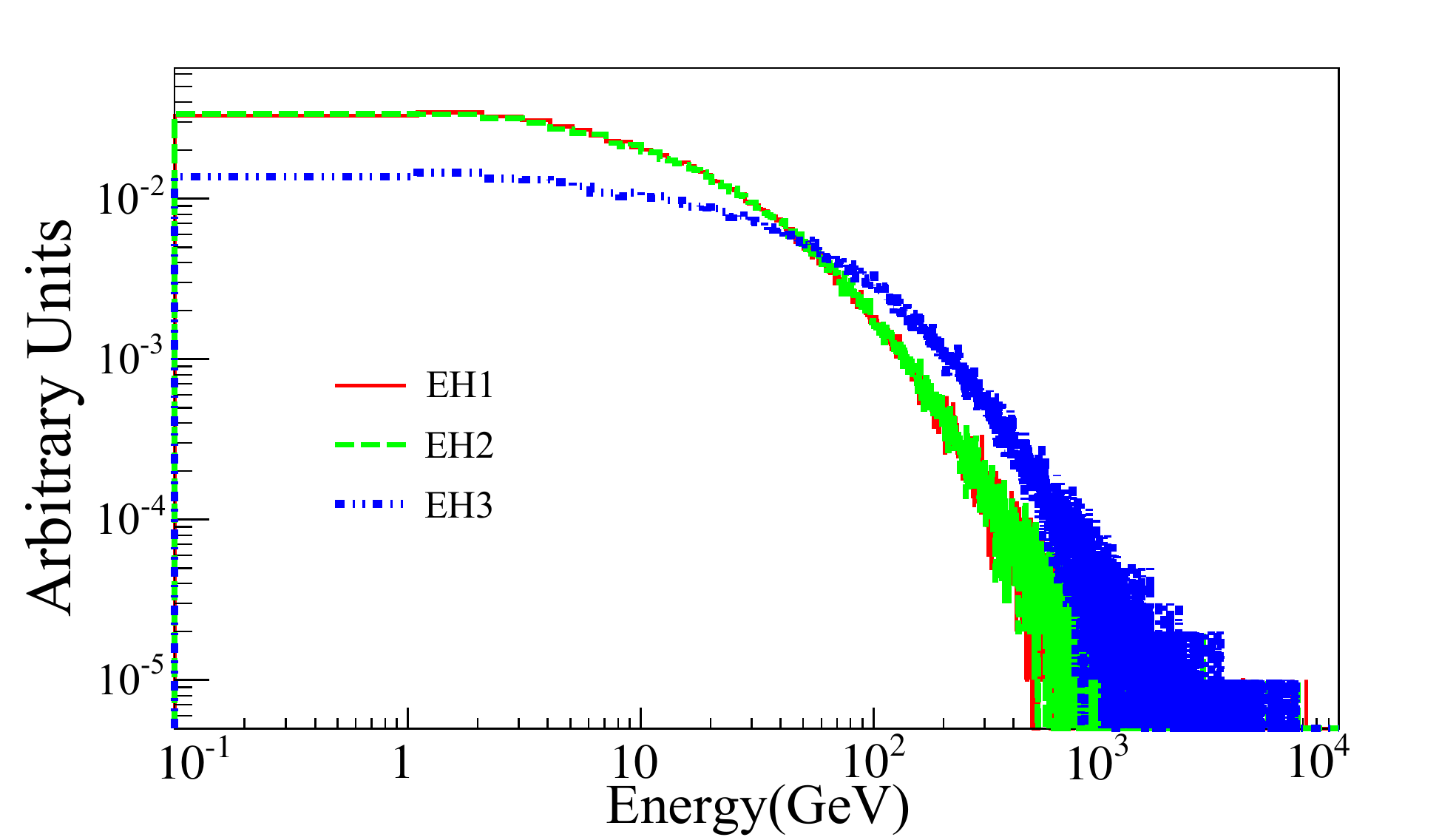}
    \caption{
       Simulated muon energy spectra.
       Each curve includes $10^6$ simulated muons.
    }
    \label{Fig12}
\end{center}
\end{figure}

A sample of $10^6$ sea-level muons was generated for each hall 
with the modified Gaisser formula.
The path length through rock of each muon was calculated using an interpolation based on the digitized mountain profile.
The muons were transported with the MUSIC code~\cite{music,music2} through the rock to get the underground muon sample.
Table~\ref{tab:underground} shows the simulated muon flux and average energy at each hall.
Figures~\ref{Fig11} and \ref{Fig12} show the simulated muon angular and energy distributions at each hall.
The simulation error from MUSIC is about 1\%.
Considering the uncertainties in mountain profile mapping, rock composition,
and density profiling, the total error in the simulated muon flux is estimated to be about 10\%. 
\begin{table}
  \centering
  \caption{
    Underground muon simulation results. The error in the simulated flux is about 10\%.
    All values have been transformed into a detector-independent spherical geometry.
  }
  \vspace{5pt}
  \begin{tabular*}{7cm}{l@{\extracolsep\fill}ccc}
    \hline
    \hline
    {\sf Hall}  & {\sf Overburden}        & {\sf Muon flux} & {\sf Average Energy} \vspace{-5pt} \\
                & {\sf m \quad mwe}       & {\sf Hz/m$^2$ } & {\sf GeV }           \\
    \hline
    EH1         & \phantom{0}93 \quad 250 &   1.27          &  57                  \\
    EH2         &           100 \quad 265 &   0.95          &  58                  \\
    EH3         &           324 \quad 860 &   0.056         & 137                  \\
    \hline
    \hline
  \end{tabular*}
  \label{tab:underground}
  \centering
\end{table}

The water shield simulation included detailed Cherenkov photon emission and propagation through the water
with conservative assumptions of water attenuation length and PMT quantum efficiency.
The propagation of Cherenkov light was modeled with the full detector geometry,
including reflection from the ADs and IWS/OWS optical barriers.
The Tyvek ~reflectivity in water was described by a full optical model in Geant4 with several parameters
which were tuned based on studies in the IHEP prototype tank.
The simulation also included the expected behavior of the electronics.
Comparisons with real data revealed that the simulation was accurate, except that some of the assumptions
were too conservative -- the water Cherenkov detector performs better than the simulations predicted.
Various scenarios of PMT failures were also modeled,
demonstrating that the pools are populated with more PMTs than needed to reach design-performance.
For example, a simulated failure of 20\% of the PMTs, randomly distributed,
was compensated for by reducing the multiplicity threshold,
thereby achieving the same muon tagging efficiency as with no failures.

The RPC system was simulated with the full detector geometry,
including the layout of the modules and the layout within a module:
four layers of RPCs, readout strips, and buffer materials, the latter corresponding to the labeled planes in Fig.~\ref{Fig5}.
The RPC array and RPC telescopes were positioned according to on-site surveys.
Individual RPCs incorporated both Bakelite and RPC gas properties.
The gas gaps were sensitive to any particle that deposited energy
and were assigned individual efficiencies determined by on-site calibration~\cite{CalRPC}.
All the electronic components mentioned in \S~\ref{SubReadout} were also simulated.
Random noise events were mixed with muon events based on a global bare RPC noise rate.
In addition, dead areas were implemented with the 1~cm-wide RPC frames and a simplified geometry for the button spacers.
Given that the module supports have a sloping angle of about $3^\circ$ (Fig.~\ref{Fig9}),
the azimuthal orientation of the supports was chosen to minimize the effective dead area in the direction of maximum muon flux.
According to simulations, this choice could affect acceptance by 0.3-0.6\%, depending on the hall. 
A comparison between simulation and data of the angular distribution of reconstructed muons is presented in \S~\ref{SubMuonDist}.

 \section{Performance}
\label{SubPerformance}
From December 2011 through July 2012, the Daya Bay experiment collected data with six 
ADs installed: two ADs in EH1 (AD1 and AD2), one in EH2 (AD3), and three in EH3 (AD4, AD5, and AD6).
By September 2012 the two final ADs were installed.
This section discusses data taken in the first period, when only six of the eight ADs were deployed.
 \subsection{PMT gains and detector rates}
\begin{figure}
\begin{center}
    \includegraphics[width=7.8cm]{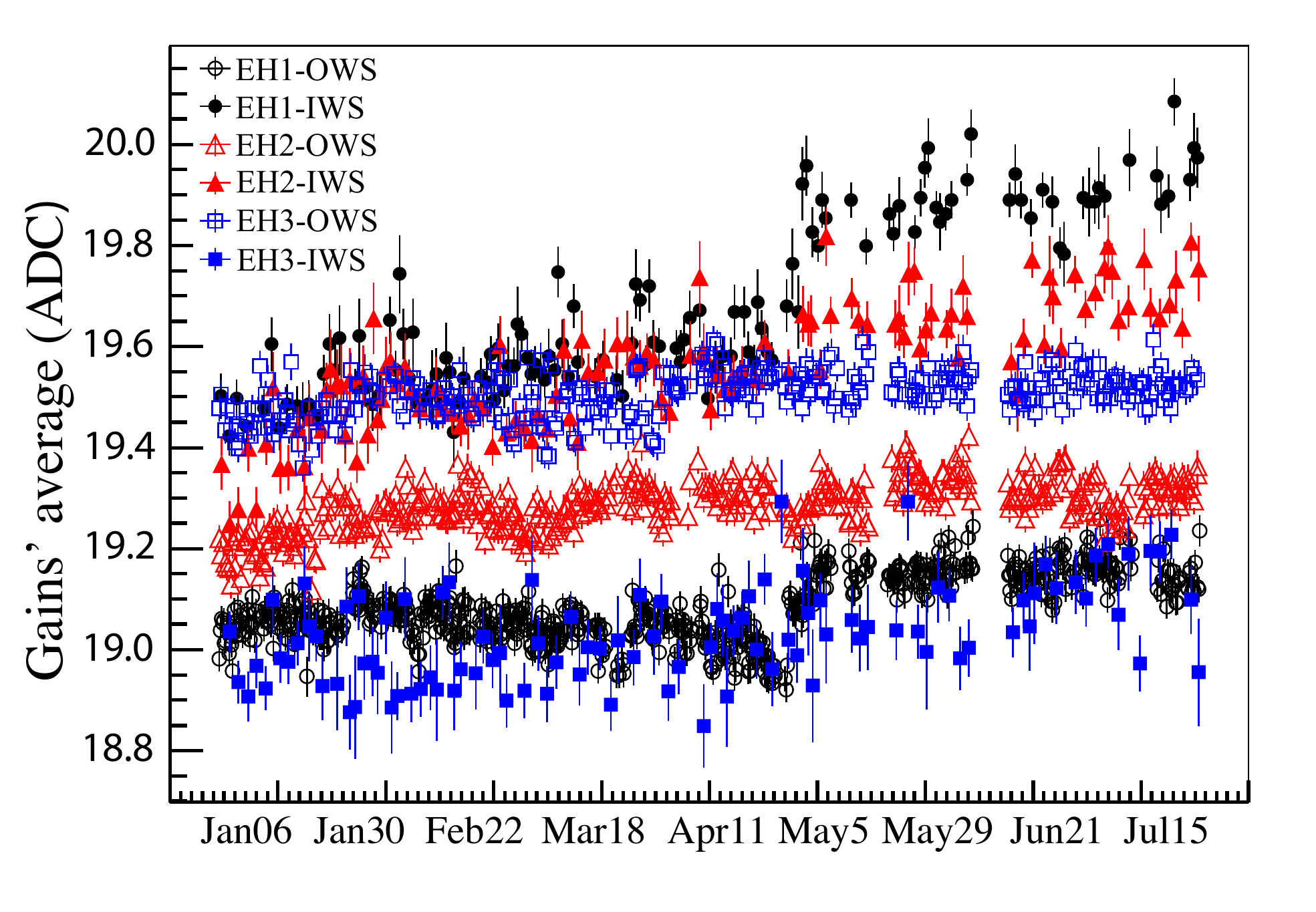}
    \caption{ Water Cherenkov PMT gains (single photoelectron charge) vs. time. 
              The jumps in the EH1 gains in early May were caused by temperature changes in the electronics crates.
              There is about three times more activity, and therefore more frequent calibration,
              in a given OWS than in the corresponding IWS.
            }
    \label{Fig13}
\end{center}
\end{figure}

\begin{figure}
\begin{center}
    \includegraphics[width=7.8cm]{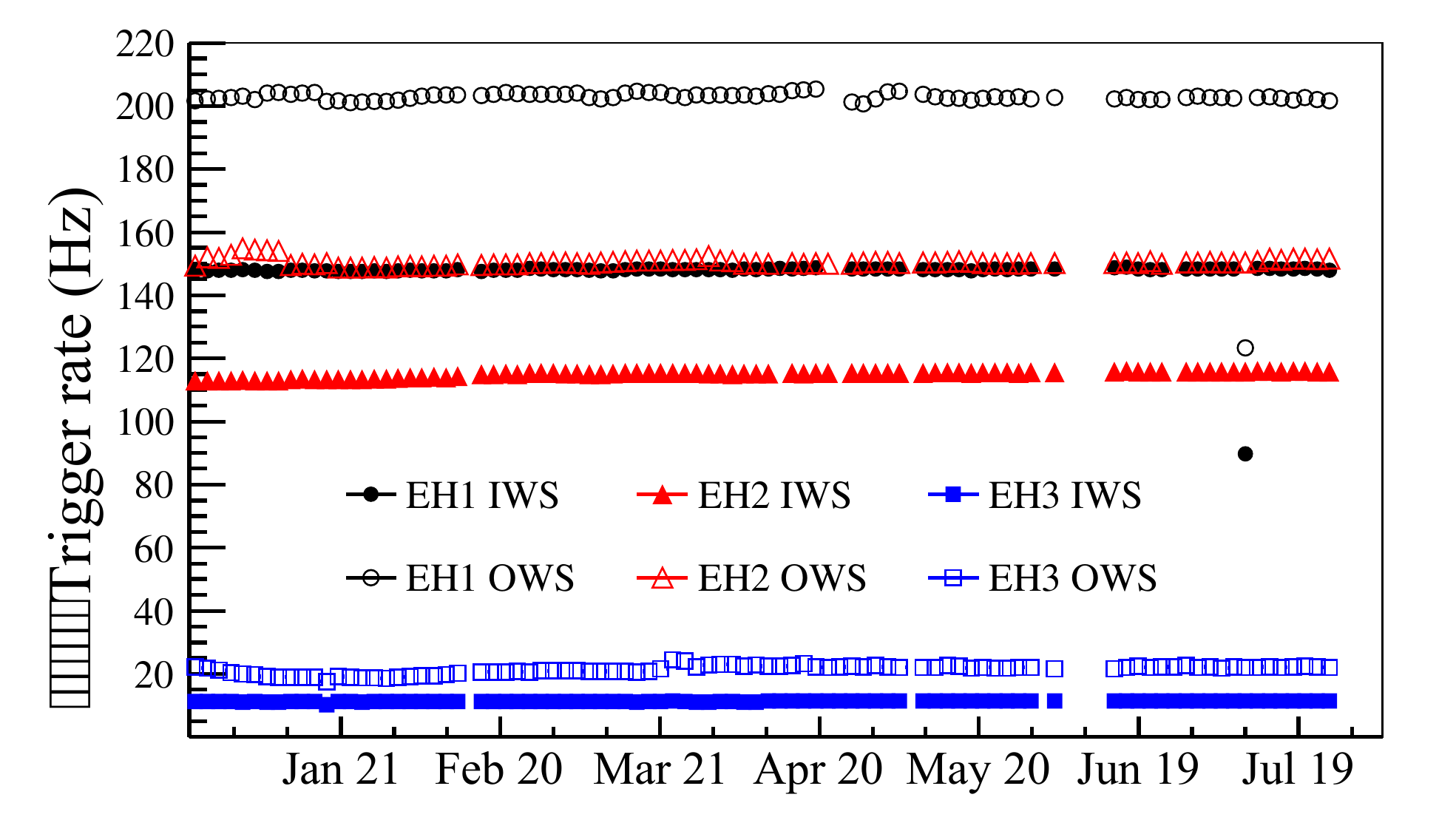}
    \includegraphics[width=7.8cm]{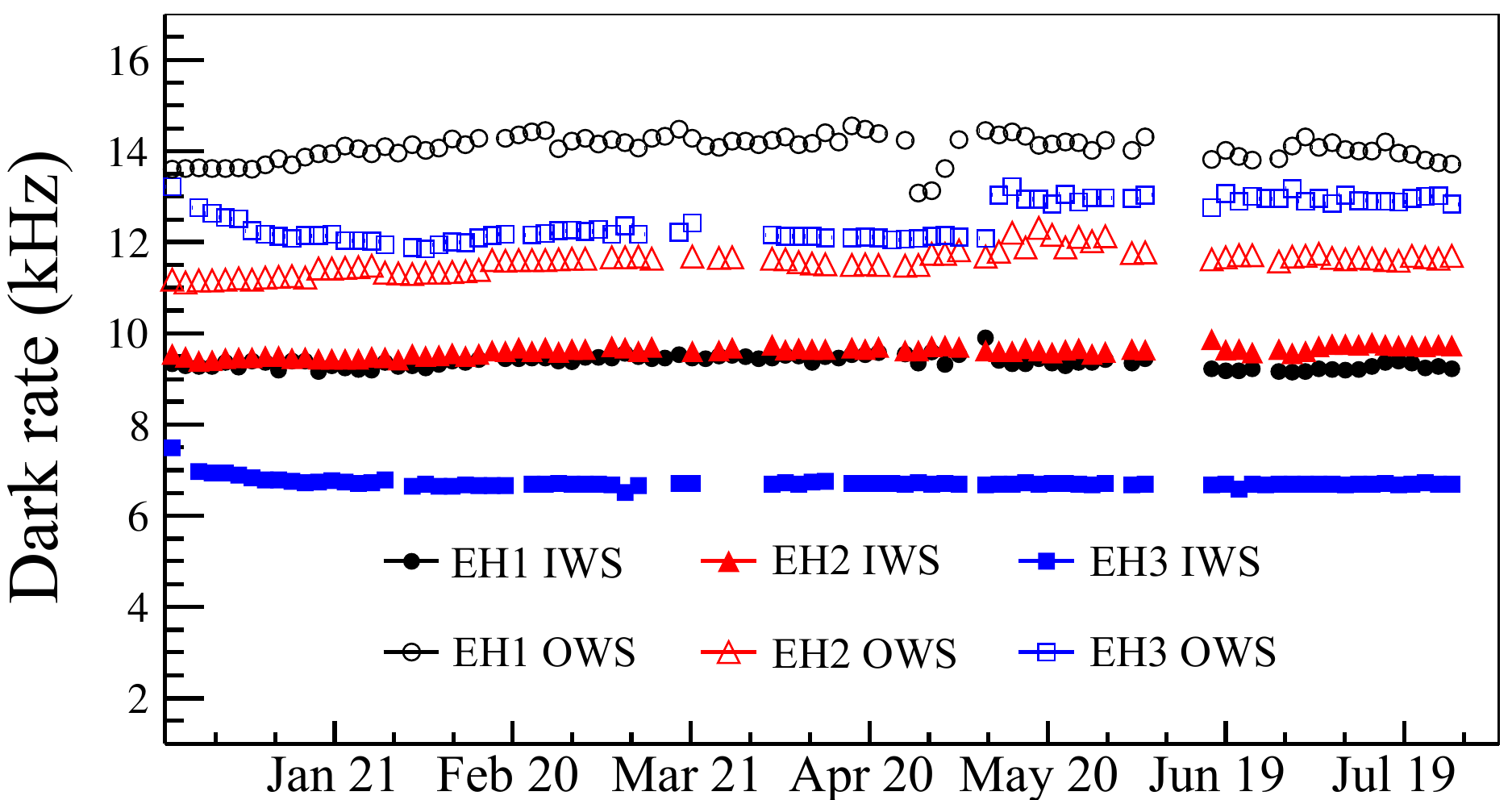}
    \caption{
       {\bf Top:} Multiplicity trigger rate vs time.
       {\bf Bottom:} Water Cherenkov PMT dark noise rate vs time, from periodic triggers.
       The gaps differ somewhat in the two plots because they are made from different trigger types.
    }
    \label{Fig14}
\end{center}
\end{figure}

Data from the periodic trigger is used to perform continuous PMT gain calibration from dark noise.
Figure \ref{Fig13} shows the water Cherenkov PMTs' average gain versus time for the three halls.
The average PMT gains are relatively stable with a slight upward trend over time.
PMT noise affects the multiplicity trigger and, therefore, the vetoing efficiency.
Figure~\ref{Fig14} shows the multiplicity trigger and average PMT dark noise rates versus time.
These do not change much over time except for a few episodes caused by noisy electronics.

\begin{figure}
\begin{center}
    \includegraphics[width=7.8cm]{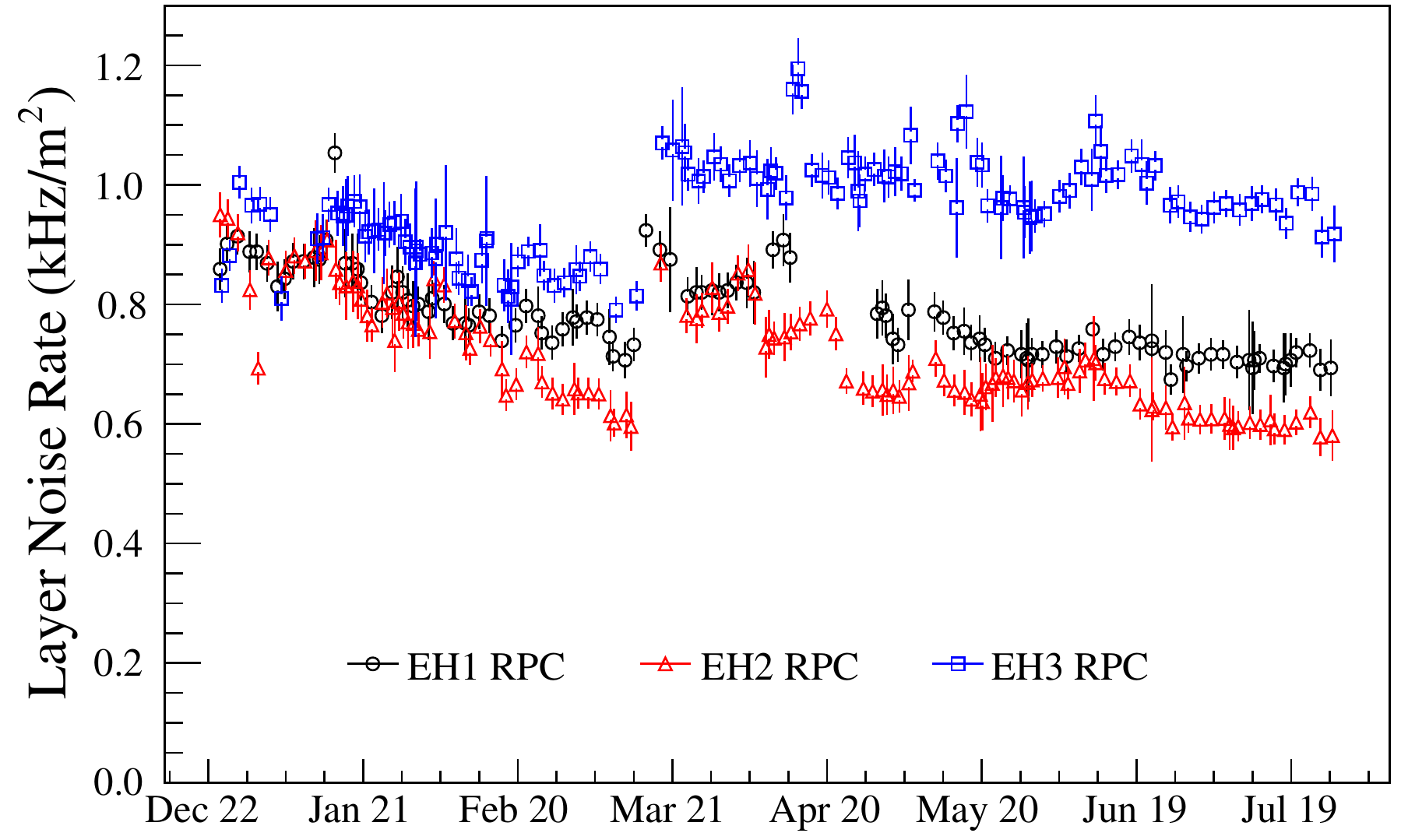}
    \caption{
      Noise rate for the RPC layers in each hall.
      On March 16, the HV was increased from 7.4 to 7.6 kV (the nominal operating voltage),
      which yielded a few percent increase in muon efficiency but also increased the noise.
      The thresholds were set to 35~mV throughout this period.
      Larger error bars are due to fewer samplings (shorter runs).
      The two-week gap in EH1 in April is due to a failed mass flow controller. 
      Fine structures are due to variations in hall temperature and humidity.
    }
    \label{Fig15}
\end{center}
\end{figure}

Data from the RPC periodic trigger is used to determine RPC layer noise rates for each run. 
Figure~\ref{Fig15} shows the noise rate of the RPC layers, as a function of time.
The average rates decrease over time, as expected.

\subsection{Water quality}
\label{SubWaterQuality}
\begin{figure}
\begin{center}
    \includegraphics[width=7.8cm]{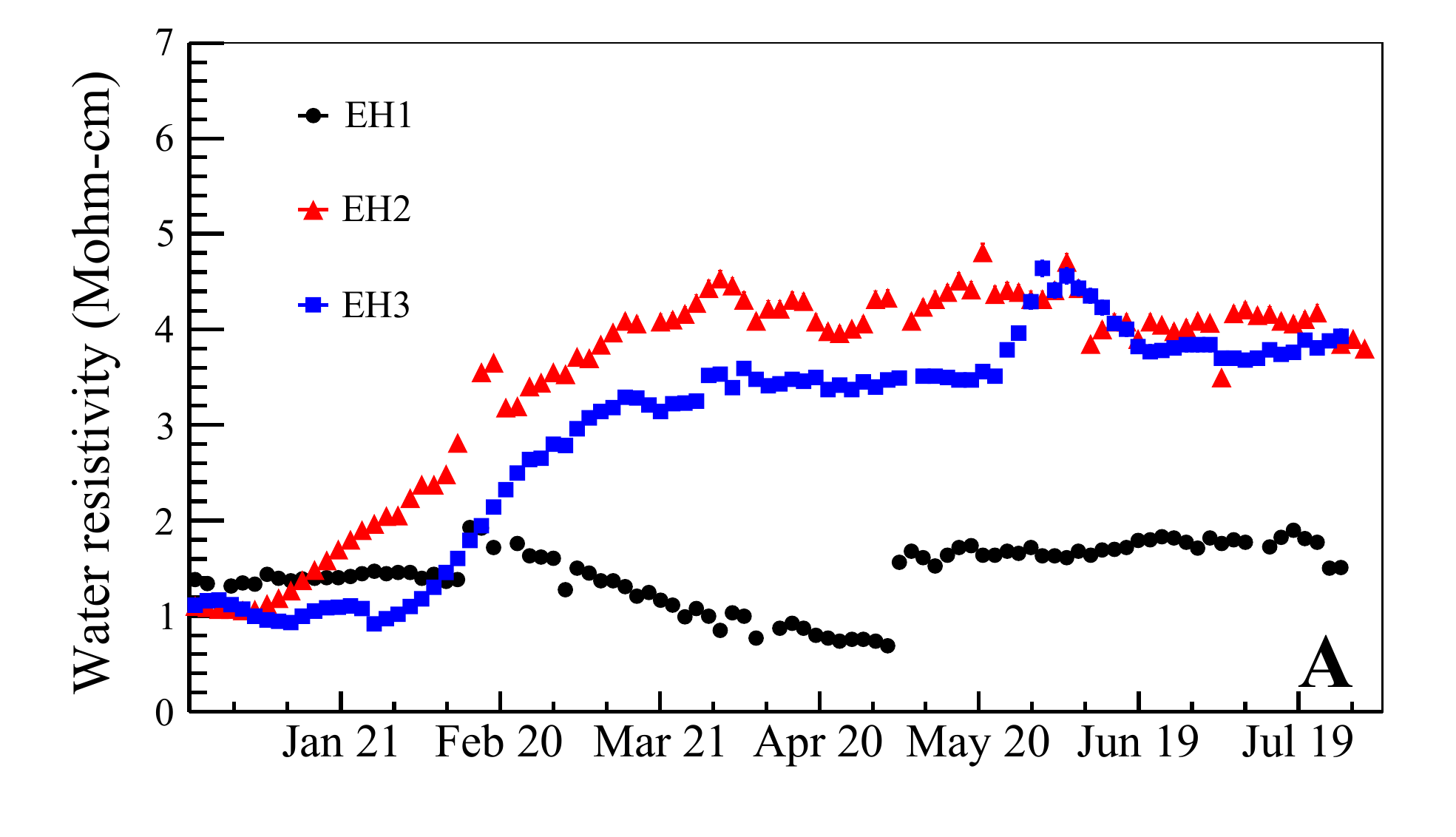}
    \includegraphics[width=7.8cm]{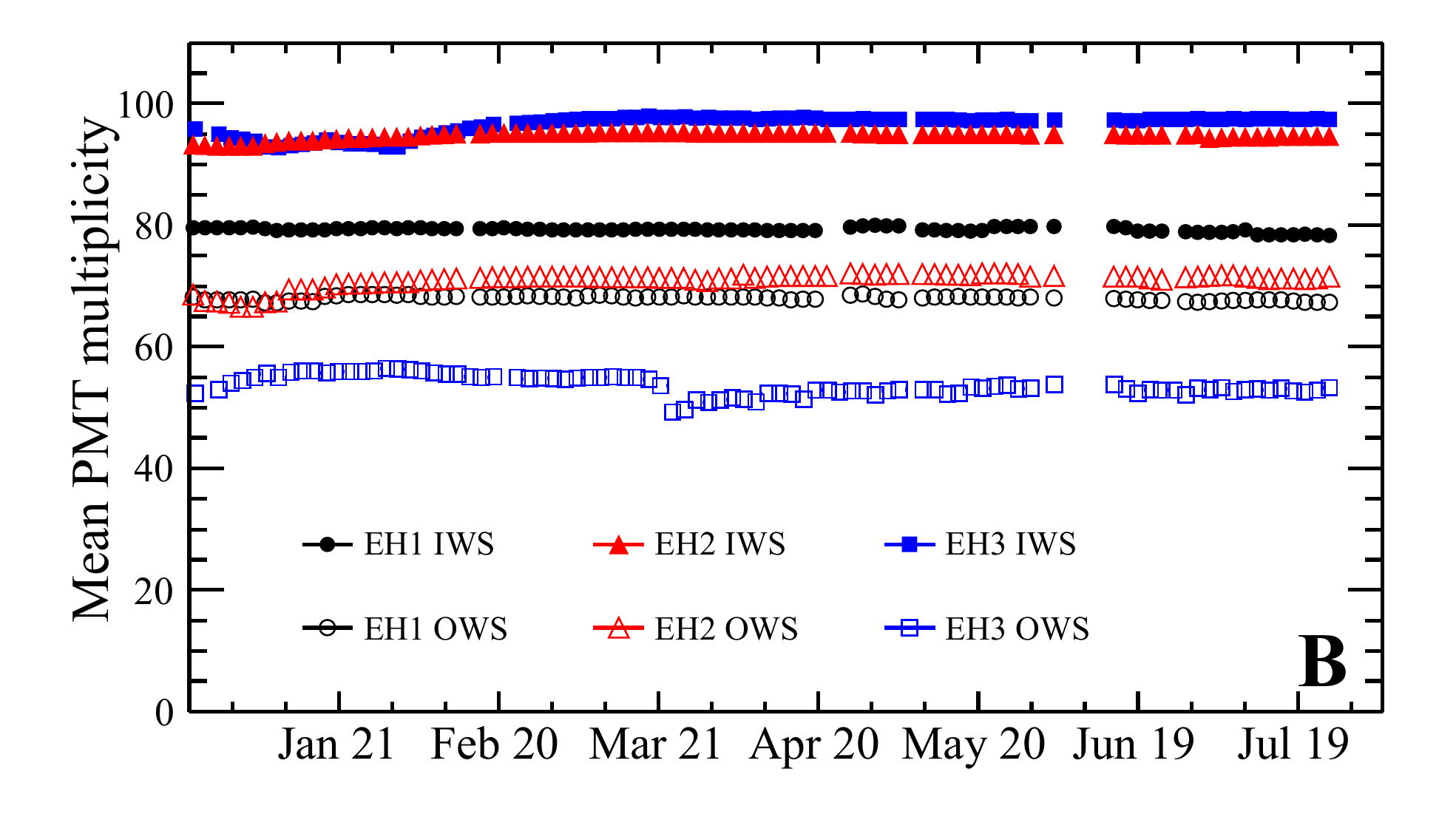}
    \includegraphics[width=7.8cm]{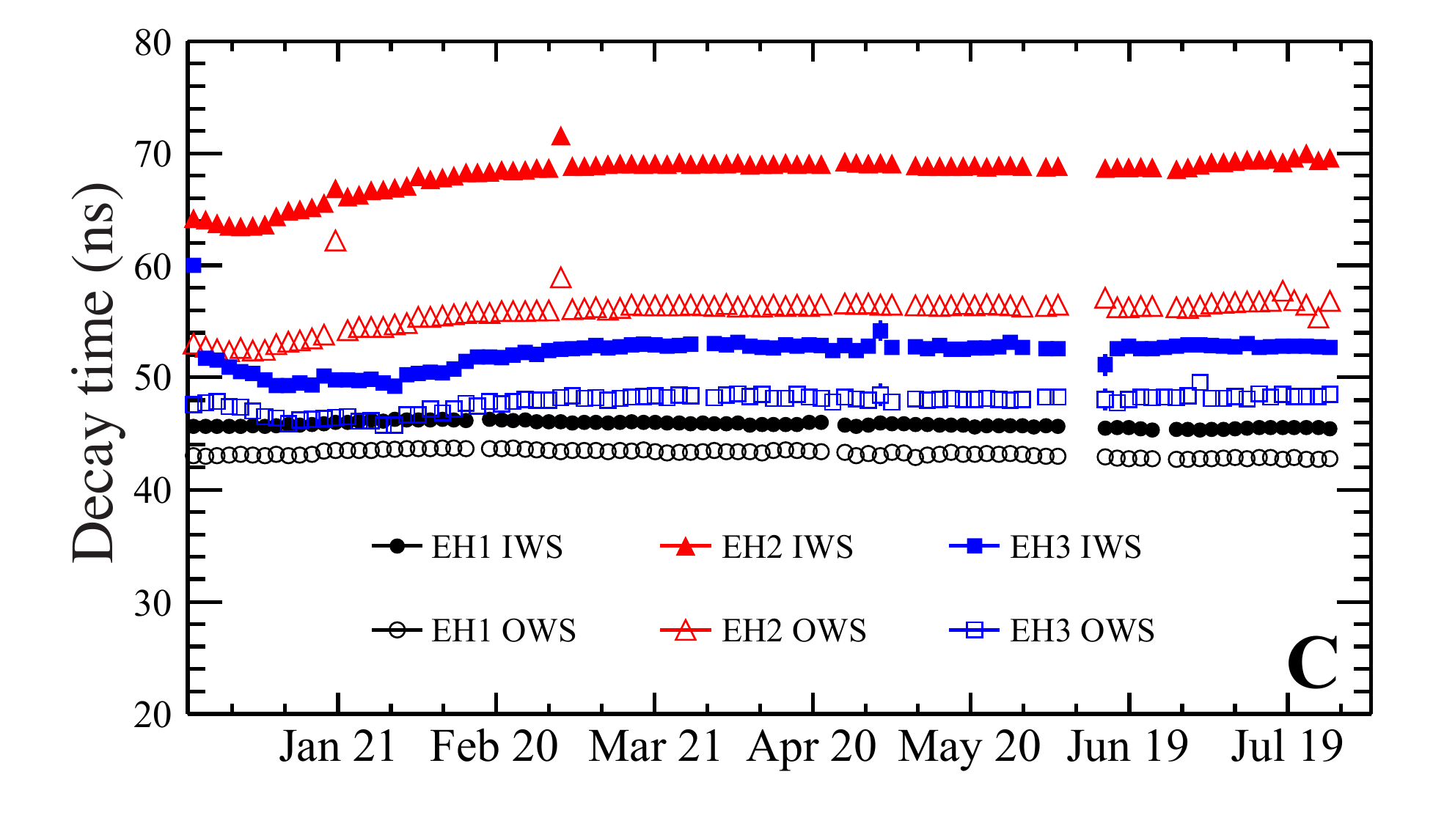}
    \caption{ Three water-clarity proxies.
              {\bf A:} Resistivity of the pool water in each hall vs. time.
              This measurement is made where water from the pool re-enters the polishing loop.
              The water delivered to the pool is typically 14 to 16~M$\Omega$-cm.
              The jump in EH1 on May 4 was due to sensor maintenance.
              {\bf B:} Mean PMT multiplicity per muon vs. time.
              {\bf C:} The decay time $\tau$ from Eq. \eqref{fit_tau},
              fit to data from short periods of time in all halls, as a function of time.
              Longer decay times correspond to clearer water (see text).
    }
    \label{Fig16}
\end{center}
\end{figure}

A readily measured indicator of water quality is the resistivity:
The higher the resistivity, the purer the water~\footnote{For ultrapure water, the maximum is 18.2~M$\Omega$-cm at 25~C.},
although this is a poor proxy for the attenuation length (e.g., adding salt to water does not change its attenuation length).
The water resistivity is continually monitored at Daya Bay (Fig.~\ref{Fig16}A).
The resistivity in EH2 and EH3 steadily increases (improves) over several weeks of operation.
In contrast, the resistivity in EH1 shows no significant improvement with time,
most likely due to an unidentified source of contamination.
This has not negatively affected the performance of the veto, however, as demonstrated below.

A better proxy for the water attenuation length is the mean multiplicity per muon~\footnote{
  Here, ``muon'' indicates that the observed PMT multiplicity exceeded 12.
  It does not indicate an actual particle identification,
  though with so many PMTs seeing light at the same time, and being deep underground, it is almost always indeed a muon.
},
plotted over a period of time in Fig.~\ref{Fig16}B.
The OWS mean multiplicities of EH1 and EH2 are very similar as might be
expected, but that of the IWS for EH2 is much higher than for EH1.
This is because during the period shown, EH2 had only one AD in the pool, whereas EH1 had two. 
That the IWS mean multiplicities of EH2 and EH3 are nearly the same is essentially accidental,
considering the larger IWS, lower muon flux, and higher trigger thresholds (Table~\ref{tab:pmtThresh}) in EH3.
The lower mean multiplicity in the EH3 OWS is due to PMT and electronics noise, which produce a larger fraction
of low-multiplicity triggers in the EH3 OWS than elsewhere.

\begin{figure}
\begin{center}
    \includegraphics[width=7.8cm]{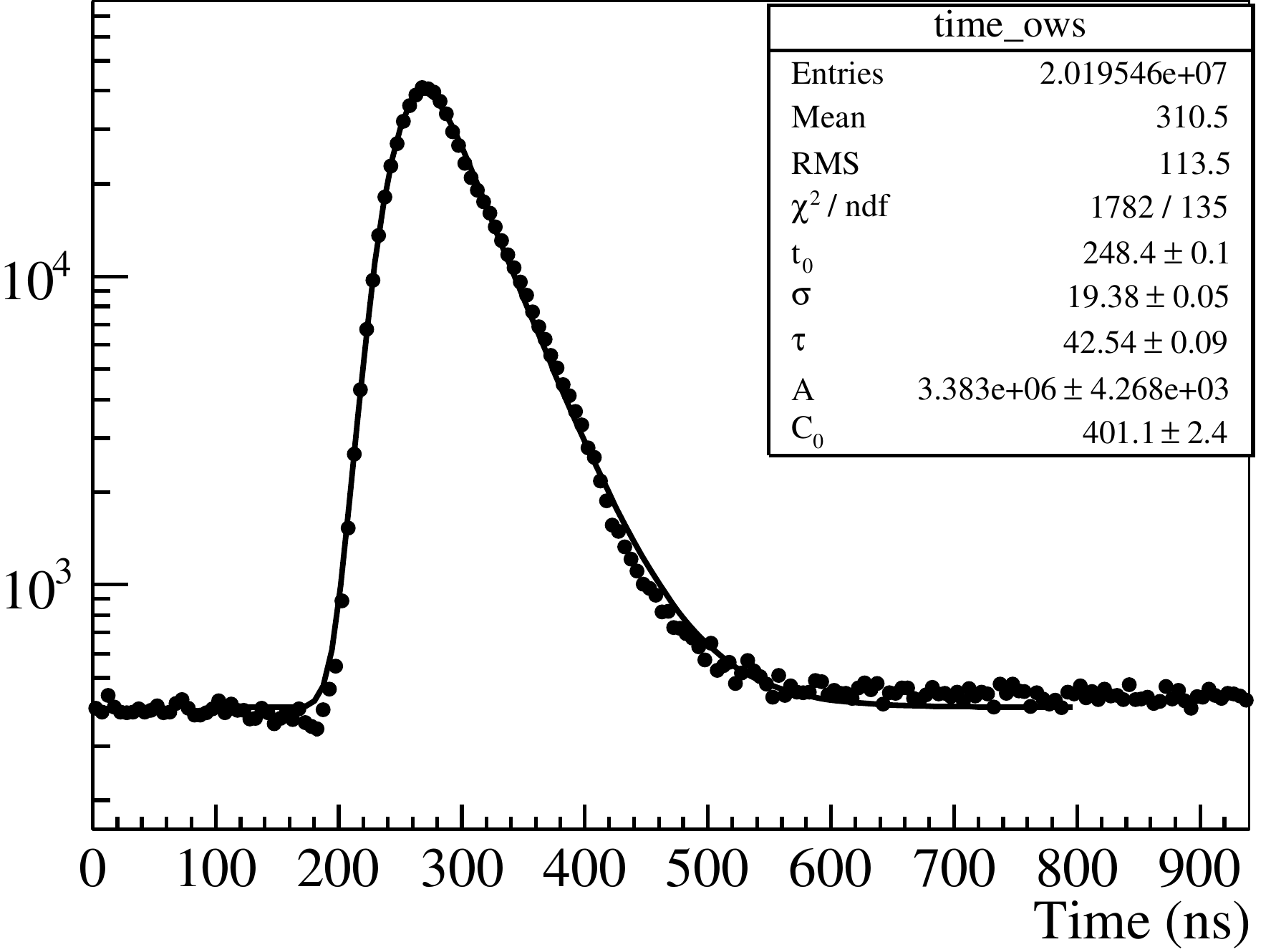}
    \caption{A typical fit to Eq. \eqref{fit_tau} to data from a short period of time in EH1 OWS.}
    \label{Fig17}
\end{center}
\end{figure}

Another proxy for the water attenuation length is the decay time of light (including reflected light)
in a pool from the passage of a muon.
This is complicated by reflection losses from the Tyvek and other materials,
some of which have a reflectivity of less than 50\% (such as the ADs),
but changes in the water attenuation length would be apparent when the decay time is followed over time,
all else remaining fixed.
To characterize this decay time in the pools,
all PMTs' threshold-crossing times are accumulated in one histogram for the IWS, and another for the OWS (Fig. \ref{Fig17}).
These two distributions are then fit with the heuristic form

\begin{align}
  \label{fit_tau}
  f(t) &= A \, b(t) \, \frac{ a_0 - a_1(t) }{ 2 \tau } + C_0 \\ \nonumber
  a_0 &= \erf \left( \frac{ \sigma^2 + \tau t_0 }{ \sqrt{2} \, \tau\sigma } \right) \\ \nonumber
  a_1(t) &= \erf \left( \frac{ \sigma^2 - \tau \left( t - t_0 \right) }{ \sqrt{2} \, \tau\sigma } \right) \\ \nonumber
  b(t) &= \exp \left( \frac{ \sigma^2 - 2 \tau ( t - t_0 ) }{ 2 \tau^2 } \right)
\end{align}
where $A, C_0, t_0, \sigma$, and $\tau$ (the light decay time) are fit parameters
and $t$ is the PMT time from the histograms.
Figure~\ref{Fig17} shows the results of fitting Eq.~\eqref{fit_tau} to some EH1 OWS data,
for which $\tau = 42.54 \pm 0.09$~ns.
The variation of $\tau$ with time for all pools is shown in Fig.~\ref{Fig16}C.
As with the mean multiplicity per muon, Fig.~\ref{Fig16}B,
the variation of the decay time $\tau$ with time is very small after the water systems had been running for about a month.
The three plots in Fig.~\ref{Fig16} indicate a slight improvement in water quality over the period covered.

\begin{figure}
\begin{center}
    \includegraphics[width=7.8cm]{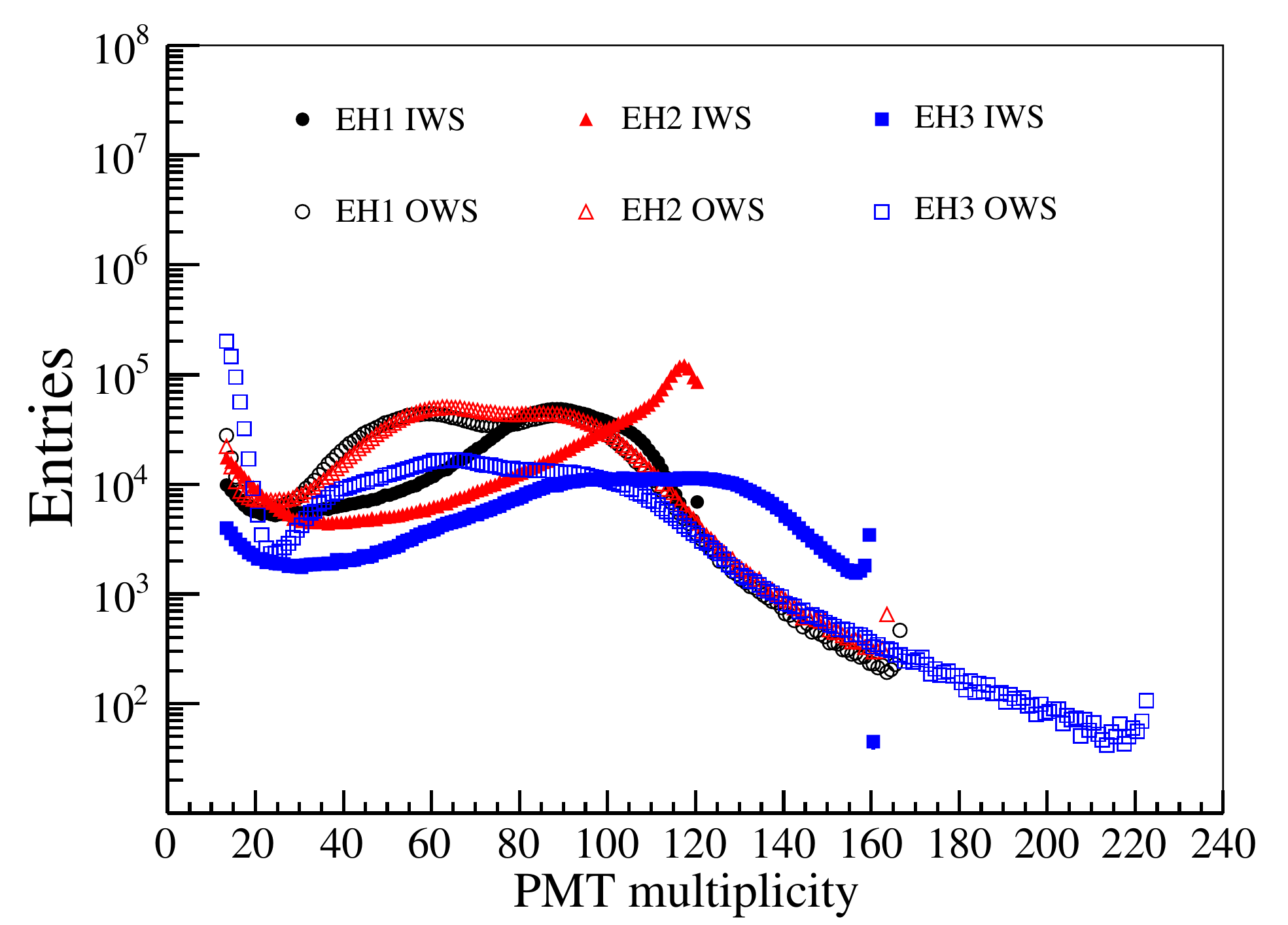}
    \caption{ Distributions of PMT Multiplicity-per-muon.
             The peak below 20 PMTs, most prominent for EH3 OWS,
             is mainly due to PMT and electronics noise.
    }
    \label{Fig18}
\end{center}
\end{figure}
Figure~\ref{Fig18} shows distributions of the PMT multiplicity per muon.
The IWS multiplicity differs from hall to hall because of the different configurations (numbers of deployed ADs).
The OWS configurations in EH1 and EH2 are the same, however, so their respective distributions are also nearly the same.
The EH2 OWS average is slightly higher than that of EH1,
which can also be seen in Fig.~\ref{Fig16}B, indicating that the EH2 water is clearer than that of EH1.

The water is essentially free of all radioactive sources, except for radon.
Radon is pervasive in an underground, granite environment, and will enter the water through any contact with
air, and through the pool walls and from all the steel in the pool as a product of the uranium or thorium decay chain.
The former is controlled by avoiding contact with air, which is one function of the pool cover gas
(it also prevents O${}_2$ and CO${}_2$ from entering the water).
The latter source is controlled by the PermaFlex, but this path of entry cannot be completely eliminated.
Along with other gasses, the water polishing stations remove radon, but it is constantly replenished.
Table~\ref{tab_radon} shows the radon activity measured in the water polishing stations.
These levels are within the Daya Bay design requirements.
\begin{table}
  \centering
  \caption{Radon activity in the water before and after polishing.}
  \vspace{5pt}
  \begin{tabular*}{7cm}{l@{\extracolsep\fill}cc}
    \hline
    \hline
    Hall & {\sf Before polishing} & {\sf After polishing} \vspace{-5pt} \\
         & {\sf Bq/m$^3$}         & {\sf Bq/m$^3$}        \\
    \hline
    EH1  & 34.7 $\pm$ 4.2         & 30.2 $\pm$ 3.6        \\
    EH2  & 86.4 $\pm$ 9.6         & 49.6 $\pm$ 5.5        \\
    EH3  & 48.4 $\pm$ 6.3         & 43.3 $\pm$ 4.9        \\
    \hline
    \hline
  \end{tabular*}
  \label{tab_radon}
  \centering
\end{table}

\subsection{Muon-detection efficiency}
\begin{figure}
\begin{center}
    \includegraphics[width=7.8cm]{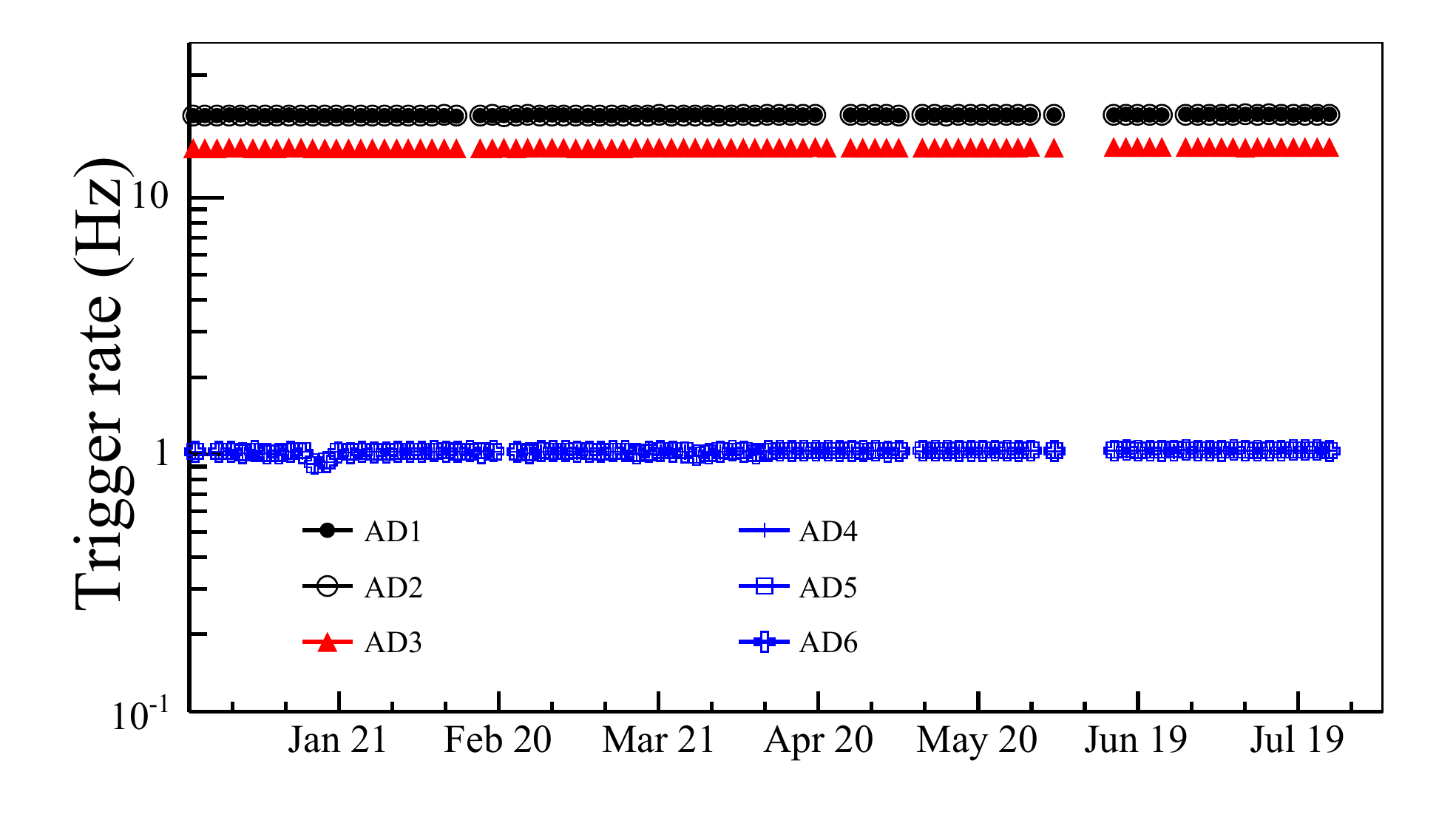}
    \caption{
      Rate of muons (depositing more than $20$~MeV in an AD) in the ADs versus time.
      Half the symbols in the plot are mostly obscured because
      the ADs' rates from a given hall are essentially the same.
    }
    \label{Fig19}
\end{center}
\end{figure}

\begin{figure}
\begin{center}
    \includegraphics[width=7.8cm]{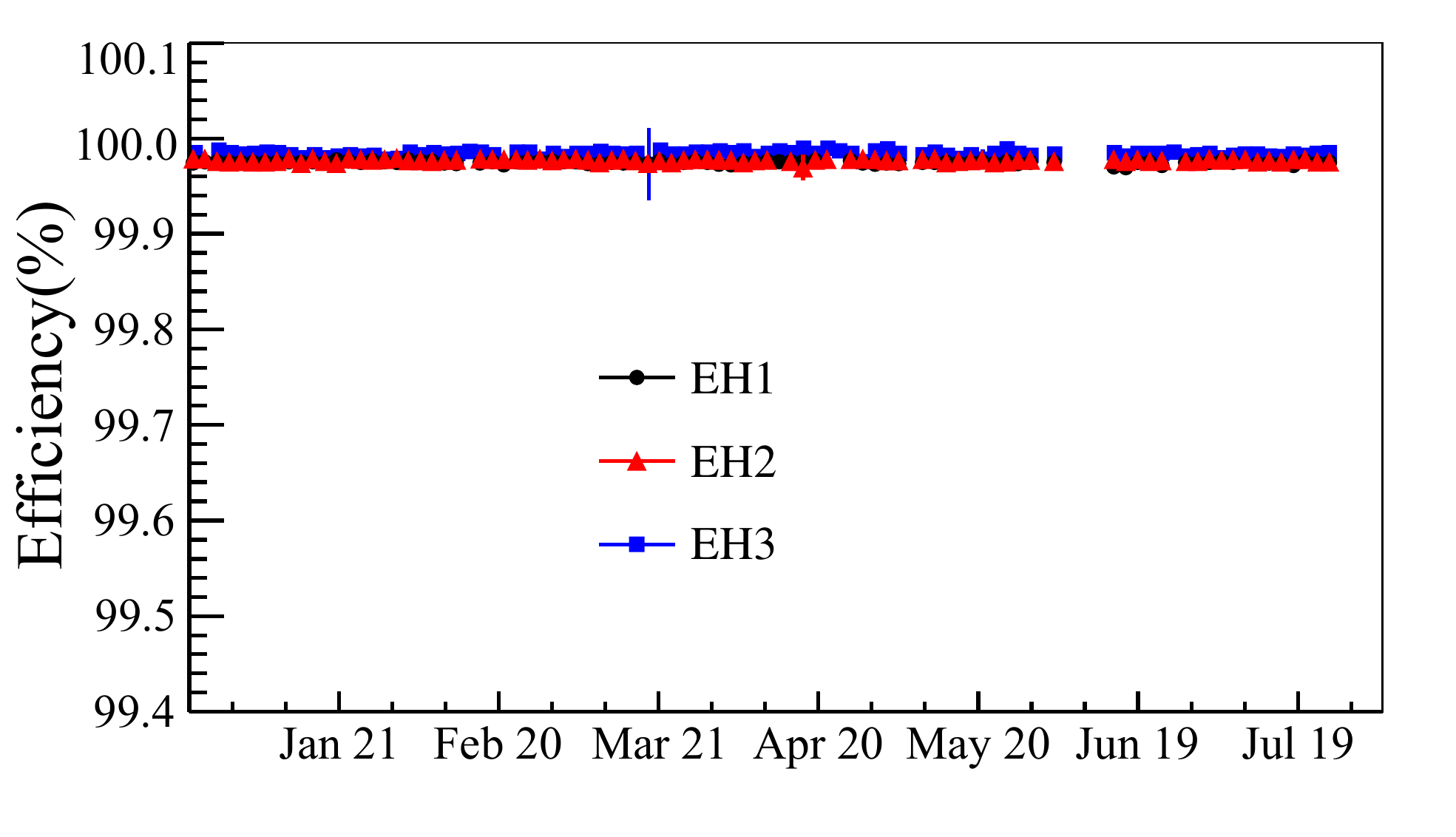}
    \caption{
      IWS efficiency for AD-tagged muons vs time.
    }
    \label{Fig20}
\end{center}
\end{figure}

Muons going through an AD (which detects them with essentially $100$\% efficiency)
can be used to measure IWS and OWS efficiencies.
Figure \ref{Fig19} gives the rate of muons in the ADs (events with more than 20 MeV in an AD).
Each AD has a muon rate of $\sim 20$~Hz in EH1, $\sim 15$~Hz in EH2, and $\sim 1$~Hz in EH3,
which correspond to the overburdens in the respective halls.
Figure \ref{Fig20} shows the IWS efficiency, for an IWS trigger of $>$12 coincident PMTs,
measured by AD-tagged muons versus time.
The mean IWS efficiency for all three halls is well-described by a single value

\begin{equation}
\label{effIWS}
\xi_{\mbox{\tiny IWS}} = 99.98 \pm 0.01\% \, .
\end{equation}
The small apparent inefficiency is due to events identified as muons in an AD,
but which are actually neutrons created by muons in the rock.
Neutrons can propagate undetected through the water into an AD,
and deposit sufficient energy to be tagged as a muon.
While simulations indicate that the IWS efficiency for AD-tagged muons should be 100\%,
they also show a very small fraction of muons showering in the rock,
where a neutron deposits energy in an AD but without a charged particle passing through the water.
This simulated fraction is consistent with the observed IWS inefficiency.

\begin{figure}
\begin{center}
    \includegraphics[width=7.8cm]{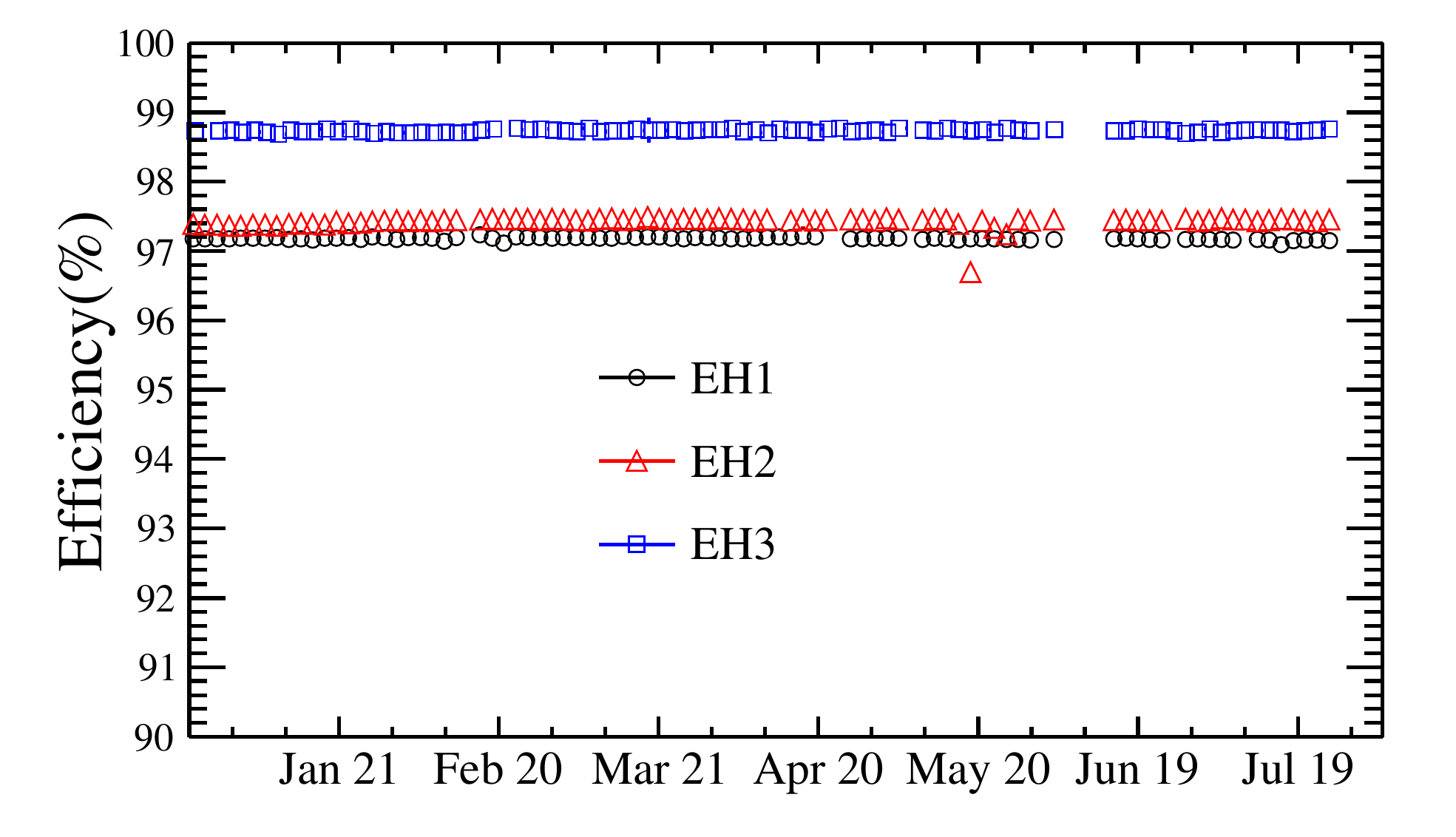}
    \caption{
      OWS efficiency for AD-tagged muons vs time.
    }
    \label{Fig21}
\end{center}
\end{figure}

Figure \ref{Fig21} shows the OWS efficiency measured by AD-tagged muons versus time.
This is expected to be lower than the true OWS efficiency because of muons that deposit energy
in the ADs but stop there or in the IWS either without traversing the OWS at all,
or by traversing only a short path in the OWS.
(Recall that the OWS does not cover the top of the IWS.)
This happens less often in EH3 because of the higher mean energy of the muons in EH3,
as can be seen by its higher AD-measured efficiency.
Muons entering through the side will be seen in the OWS and do not contribute to this inefficiency.
Table~\ref{tab:OWS} shows the measured and simulated OWS efficiencies.
\begin{table}
  \centering
  \caption{OWS measured and simulated (MC) efficiencies $\xi_{\mbox{\tiny ows}}$.}
  \label{tab:OWS}
  \vspace{5pt}
  \begin{tabular*}{7cm}{l@{\extracolsep\fill}ccc}
    \hline
    \hline
               & {\sf EH1 }         & {\sf EH2}          & {\sf EH3}   \\
    \hline
    Data       & $97.2 \pm 0.2$\%   &   $97.4\pm0.2$\%   & $98.7\pm0.2$\%  \\
    MC         & $97.4 \pm 0.1$\%   &   $96.6\pm0.2$\%   & $98.0\pm0.2$\%  \\
    Difference & $-0.2 \pm 0.2$\%   &    $0.8\pm0.3$\%   &  $0.7\pm0.3$\%   \\
    \hline
    \hline
  \end{tabular*}
  \centering
\end{table}

Table~\ref{tab:rates} shows the underground muon flux at each hall, as determined by each detector subsystem.
The selection criteria in the three systems are deposited energy $>$20~MeV for the ADs,
more than 20 PMTs for the IWS and OWS above 0.25~pe,
and four out of four triggered planes plus a strip cut for the RPCs.\footnote{
  RPC efficiency and muon flux are determined in~\cite{CalRPC}.
  The strip cut requires that each triggered layer has only one or two adjacent triggered strips.
}
Corrections are made for the efficiency of each detector.
All values are consistent with the simulation results in Table~\ref{tab:underground}.

\begin{table}
  \centering
  \caption{Measured underground muon flux (Hz/m${}^2$).
           The simulated flux from Table~\ref{tab:underground} is included for comparison.
           All values have been transformed into a detector-independent spherical geometry.
           The errors are dominated by the systematic uncertainty of the transformation,
           which is the difference between assuming the simulated vs. measured angular spectrum.
          }
  \label{tab:rates}
  \vspace{5pt}
  \begin{tabular*}{7cm}{l@{\extracolsep\fill}ccc}
    \hline
    \hline
    {\sf ~Detector~}  & {\sf EH1 }    & {\sf EH2 }    & {\sf EH3 }      \\
    \hline
    AD                & $1.21\pm0.12$ & $0.87\pm0.09$ & $0.056\pm0.006$ \\
    IWS               & $1.15\pm0.12$ & $0.86\pm0.09$ & $0.055\pm0.005$ \\
    OWS               & $1.12\pm0.11$ & $0.84\pm0.08$ & $0.053\pm0.005$ \\
    RPC               & $1.17\pm0.09$ & $0.87\pm0.11$ & $0.053\pm0.006$ \\
    Average           & $1.16\pm0.11$ & $0.86\pm0.09$ & $0.054\pm0.006$ \\
    Simulation        & $1.27\pm0.13$ & $0.95\pm0.10$ & $0.056\pm0.006$ \\
    \hline
    \hline
  \end{tabular*}
  \centering
\end{table}

\subsection{Muon angular distributions}
\label{SubMuonDist}
Muon trajectories can be reconstructed from the main RPC array and the telescope RPCs,
which are described at the end of \S~\ref{subCandI}.
The position resolution in these devices is about 10~cm which corresponds to roughly $2\degree$ in zenith
and $4\degree$ in azimuth.
This reconstruction uses the same event selection as the muon flux calculation.

\begin{figure}
\begin{center}
    \includegraphics[width=7.8cm]{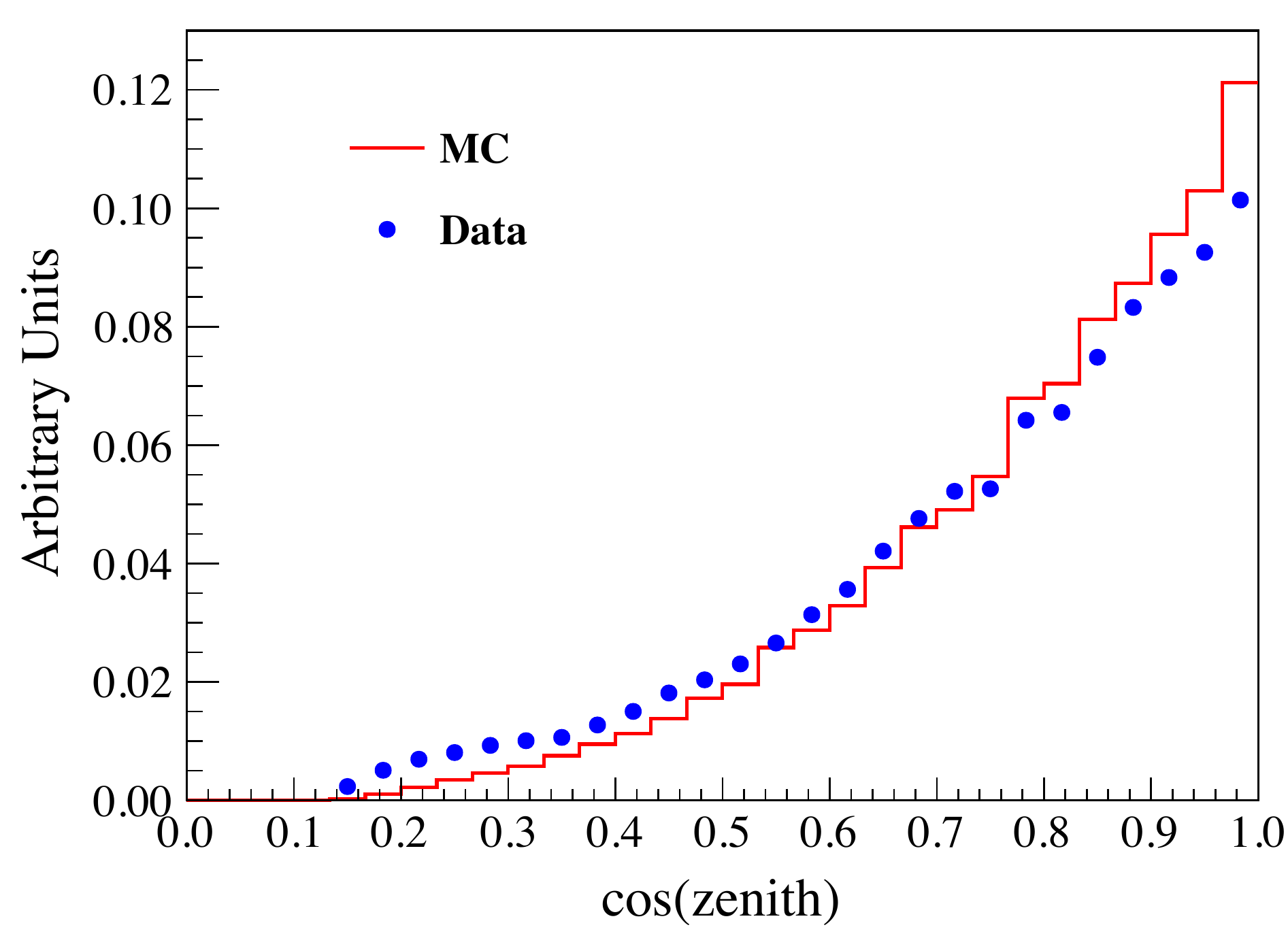}
    \includegraphics[width=7.8cm]{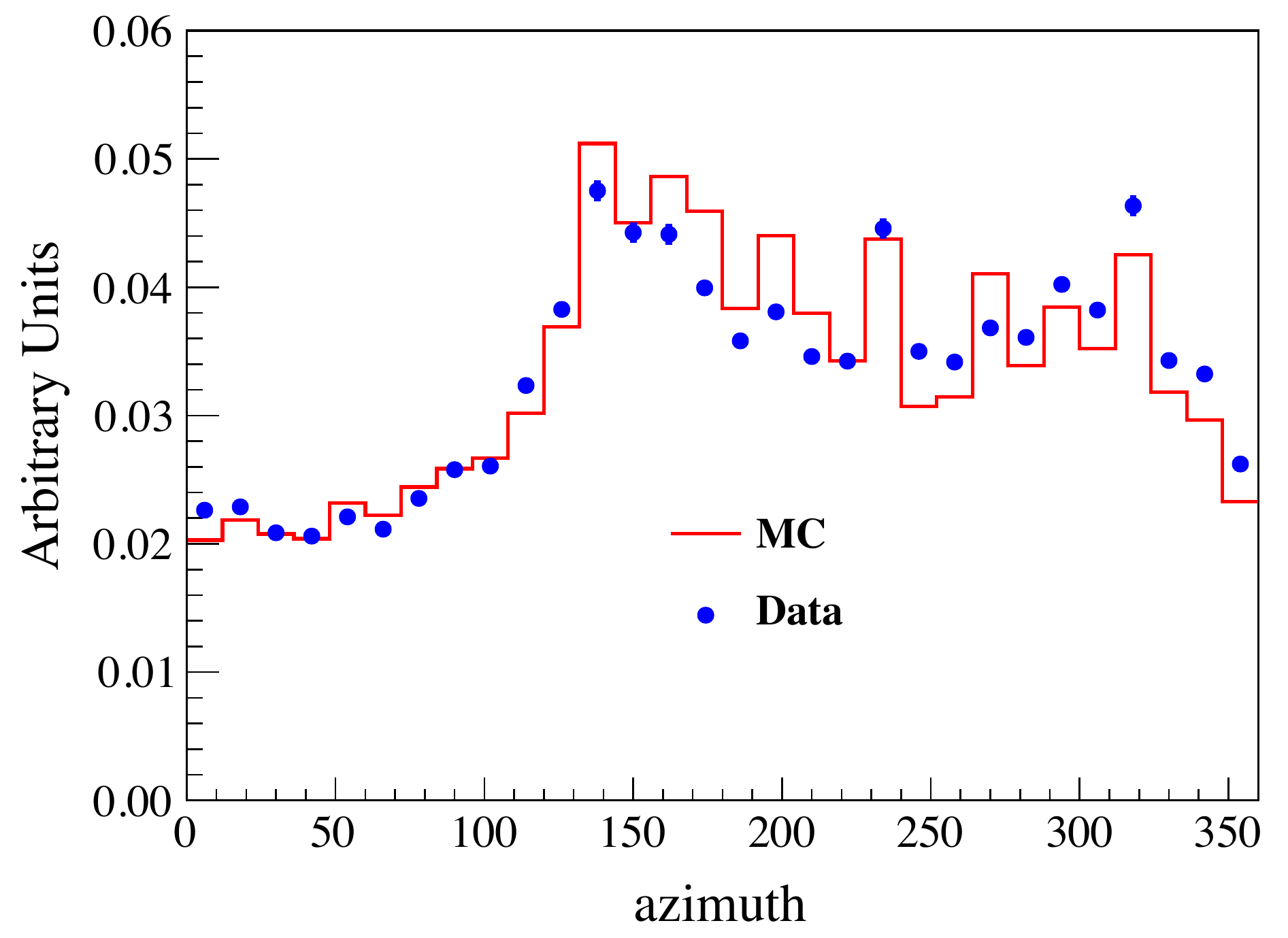}
    \caption{
      RPC-reconstructed muon trajectories in EH1.
      The MC and data points are normalized by their respective total counts.
      Corrections are made for the measured efficiencies.
    }
    \label{Fig22}
\end{center}
\end{figure}

The angular coverage of the telescopes is shown in
Fig.~\ref{Fig22}, which compares data with a simulation for EH1.
They are consistent within the reported errors in simulation and measurement.
The data-excess at large zenith might be due to accidental muon coincidences, which were not simulated.
The accidental coincidence rate between the main RPC array and the RPC telescopes increases with zenith
as the number $N_{\mbox{\tiny geo}}$
of modules geometrically available for accidentals
increases until about 75${}^\circ$, where $N_{\mbox{\tiny geo}}$ begins to decrease.
The bare RPC noise rate is several orders of magnitude too small to make a noticeable contribution
to these accidentals (see Fig.~\ref{Fig15}).

\subsection{PMT failures}
Between the time when the EH1 pool was first filled in August of 2011 and the shutdown in the summer of 2012,
13 PMTs had failed in the three halls.
These were all EMI PMTs, recycled from the MACRO experiment~\cite{macro}.
Because the EMI PMTs were all deployed in the top two meters of the pool, these failures were
all near the top of the pool, but otherwise randomly distributed.
Most of these PMTs failed in the first month or two after their respective pool was filled.
Failures continued after this initial period, though at a reduced rate.
During the summer of 2012, most of the failed PMTs (i.e., those that were accessible)
were replaced with Hamamatsu PMTs during maintenance operations
which included the installation of the final two ADs in EH2 and EH3.
Examination of the failed PMTs revealed that they had mostly suffered very small cracks in the glass,
through which the vacuum drew in water.
Only one of these PMTs had shattered.
Upon refilling in September 2012, nine more PMTs immediately failed, all EMI.
As of August 2014, another 14 PMTs failed, with a mean time between failures of about 40 days.
Of these latest failures, essentially at random intervals of time, three were Hamamatsu PMTs (one in each of the three halls)
and the rest were EMI. 
As with the first 13 failures, these later EMI failures were all near the top of the pool, and randomly distributed.
One Hamamatsu failure was on the bottom of the pool, while the other two were about two meters above the bottom.
The total fraction of failed PMTs is just below 4\%,
far below the simulated 20\% failures which was shown to have no effect on performance.

\section{Conclusion}
The muon system performs as required by the design specifications,
with a muon detection efficiency not significantly different from 100\% for the IWS alone.
The small apparent IWS inefficiency of $0.02\pm 0.01$~\%, as determined from muons identified by the ADs,
is consistent with neutron background in the ADs.
The somewhat larger apparent inefficiency in the OWS is consistent with muons that stop in the AD or IWS without reaching the OWS,
rather than a true inefficiency in the OWS.
Indirect studies 
show that the water attenuation length is about 40~m,
but it has not yet been definitively measured.
The independent proxies for quantifying water quality discussed in \S~\ref{SubWaterQuality} support this level of clarity,
and demonstrate that the water-compatibility program mentioned in \S~\ref{SubWater} was successful.
PermaFlex has proved itself to be an effective coating for the pool:
Had the water made direct contact with the concrete, the water quality could not have been as good as observed.
The water treatment systems have also performed well.

Unplanned interventions and modifications made as needs arose improved the experiment,
such as the covering or replacement of the fluorescing insulating material on top of the AD stands,
and the addition of a dry air system around the RPCs to reduce the dark currents which were higher than expected
because of the high ambient humidity of the sub-tropical climate at Daya Bay.
Although it occurred in the (late) planning stage,
the recognition that the warm climate at Daya Bay required cooling of the water
should be considered to be in this same category of interventions,
since the original plan specified only that the water would be heated.
Measurements of the underground rock temperature revealed that the water would be too warm, rather than too cool,
reflecting a northern-climate bias in the early planning stage.
The constructed cooling capacity was successful in maintaining its design goal of 24$\pm 1$~C water.

In addition to its main function as background shield and muon veto,
the muon system also serves to study background in the IBD signal.
There remain contributions to this background from muons that completely miss the pools and the RPCs, or have only very short
paths in the pools,
while producing fast neutrons in the rock which find their way to an AD and produce a false IBD signature.
Small corrections for these contributions, based on simulations, are made in the full neutrino analysis~\cite{hep-ex/0701029}.
The results presented in \S~\ref{SubPerformance} demonstrate the veracity of these simulations.

Given the much higher failure rate of the EMI PMTs, the decision to recycle these PMTs from the MACRO
experiment was perhaps somewhat unfortunate.
Nevertheless, this has had no impact on the results of the Daya Bay experiment.
With a mean time between PMT failures of 40~days, the Daya Bay Experiment will have completed a five year run long before
20\% of the water pool PMTs have failed,
the level of failure which has been shown by simulations to have no impact on the muon veto efficiency.

\section{Acknowledgments}
The Daya Bay experiment is supported in part by
the Ministry of Science and Technology of China,
the United States Department of Energy,
the Chinese Academy of Sciences,
the National Natural Science Foundation of China,
the Guangdong provincial government,
the Shenzhen Municipal government,
the China General Nuclear Power Corporation (formerly the China Guangdong Nuclear Power Group),
Shanghai Laboratory for particle physics and cosmology,
the Research Grants Council of the Hong Kong Special Administrative Region of China,
the focused investment scheme of CUHK and University Development Fund of The University of Hong Kong,
the MOE program for Research of Excellence at National Taiwan University and NSC funding support from Taiwan,
the U.S. National Science Foundation,
the Alfred P. Sloan Foundation, 
the  University of Wisconsin,
the Virginia Polytechnic Institute and State University,
Princeton University,
California Institute of Technology,
University of California at Berkeley,
the Ministry of Education, Youth and Sports of the Czech Republic,
the Joint Institute of Nuclear Research in Dubna, Russia,
and the NSFC-RFBR joint research program.
We are grateful for the ongoing cooperation from
the China General Nuclear Power Corporation.

\bibliographystyle{elsarticle-num}
\bibliography{muonDYB}

\end{document}